\documentclass[reprint,superscriptaddress,amsmath,floatfix,aps,prd,twocolumn]{revtex4-2}

\usepackage{amssymb}  
\usepackage{graphicx}
\usepackage{exscale}
\usepackage{textcomp}
\usepackage{enumerate}
\usepackage{amsmath}
\usepackage{amssymb}
\usepackage{multirow}
\usepackage{orcidlink}

\usepackage{hyperref}
\hypersetup{breaklinks=true, colorlinks=true, citecolor=blue}
\usepackage{color}
\usepackage[normalem]{ulem}  

\begin{document}


\title{Investigating the role of nuclear parameters\\ in Neutron Star oscillations: a model comparison}
\author{Rajesh Maiti\,\orcidlink{0009-0008-5947-3060}}
\email{rajesh.maiti@iucaa.in}
\affiliation{Inter-University Centre for Astronomy and Astrophysics, Ganeshkind, Pune 411007, India}

\author{Debarati Chatterjee\,\orcidlink{0000-0002-0995-2329}}
\email{debarati@iucaa.in}
\affiliation{Inter-University Centre for Astronomy and Astrophysics, Ganeshkind, Pune 411007, India}

\begin{abstract}
Recent studies based on the relativistic mean field (RMF) model found certain nuclear empirical parameters, in particular the nucleon effective mass, to be strongly correlated with observable properties of Neutron Stars (NSs), such as the frequencies of $f-$mode oscillations. This shows the potential to constrain the values of effective mass from future observations of $f-$modes. One of our primary goals of this work is to investigate whether such correlations are physical or an artifact of the underlying nuclear model. To test this, we perform a comparative study of the correlations between NS astrophysical observables and nuclear physics parameters using two different equation of state models based on RMF theory and non-relativistic Meta-Modelling (MM) scheme. The nuclear meta-model does not assume any underlying nuclear model and therefore allows us to test the model dependence of the results. The calculations of the $f-$mode characteristics are performed within the relativistic Cowling approximation. We use state-of-the-art nuclear microscopic calculations at low density and multi-messenger astrophysical data at high-density within a Bayesian-inspired scheme to constrain the parameter space of the nuclear models. From the posterior distribution, we probe the underlying correlations among nuclear parameters and with NS observables. We find that the correlation between the symmetry energy and its slope is physical, while that of the nucleon effective mass with NS observables is model-dependent. The study shows that the effective mass governs the high density behaviour in RMF models, while in the MM it is controlled by the higher order saturation parameters, and hence probes the possibility of constraining them from future $f$-mode observations. The findings of this investigation are interesting both for astrophysics as well as nuclear physics communities.
\end{abstract}




\maketitle

\section{Introduction}
\label{sec:intro}

Since the discovery of Neutron Stars (NSs), one of the primary motivations for studying these compact objects has been to infer the properties of dense matter under extreme conditions. While the scope of studying nuclear properties with nuclear experiments is limited to densities close to saturation ($n_{sat} \sim 0.16 fm^{-3}$), heavy-ion collision experiments in particle accelerators allow us to probe dense matter properties up to $\sim 2-3 n_{sat}$. As the densities in the NS interior (up to $8-10 n_{sat}$) are much higher than those accessible by terrestrial experiments, it provides a natural astrophysical environment to probe the nature of dense nuclear matter~\cite{Glendenning_book,Schaffner-Bielich_book, Haensel_2007_book, Blaschke_2018_REV}.

The relation between the pressure and density i.e. {\it Equation of state} (EoS) is the main entity that governs the observable NS properties~\cite{Heiselberg2000,Lattimer2012Rev, Baym_2018, BURGIO2021,Lattimer_2021_REV, Sumiyoshi_2022_REV, chatziioannou_2024_REV}. The composition of the NS interior (such as presence of strangeness) may affect its EoS and hence the bulk NS properties such as its mass, radius, moment of inertia, and tidal deformability. Given the EoS, it is possible to describe the NS mass-radius relations by solving Tolman-Oppenheimer-Volkoff (TOV) equations (which ensure mass conservation and hydrostatic equilibrium) simultaneously with the EoS  \cite{Glendenning_book, Schaffner-Bielich_book}. A stiffer EoS corresponds to a large maximum mass of the NS and vice versa. Therefore, the mass and radius are the significant observables to constrain the EoS of the neutron star. Another important observable in binary NSs is the tidal deformability ($\Lambda$), which depends on the compactness ($C = M/R$) parameter of the NS \cite{Hinderer_2008_APJ,Hinderer_2010}. One can use these observables to constrain the EoS of the NS to improve the understanding of the NS interior composition.

Neutron stars are usually observed as pulsars (rotating and highly magnetized NS). NSs can be observed at multiple wavelengths across the electromagnetic spectrum with ground-based as well as space-based telescopes (e.g. Fermi, INTEGRAL, Chandra, LOFAR, GMRT etc). One can derive NS observables from multi-messenger astronomical data. In the case of neutron stars in binary, one can measure the mass to high precision from post-Keplerian parameters~\cite{Thorsett_1998, Vidana_2020}. From X-ray observations one can obtain gravitational redshift and compactness, from which one can calculate the radius of the NS given the mass. X-ray observations also provide information about the NS cooling rate. However, the precise measurement of NS radius is still a challenge.  Using advanced techniques onboard the NICER (Neutron Star Interior Composition ExploreR) mission~\cite{NICER_2014} and the planned eXTP (enhanced X-ray Timing and Polarimetry Mission)~\cite{eXTP_2018}, we expect to measure the masses and radii of NSs simultaneously with even higher precision.

Non-axisymmetric perturbations in NSs can produce Gravitational Waves (GWs)~\cite{Dhurandhar_2022_book, Riles_2022_REV}. Neutron stars are prominent GW sources in both isolated and binary systems. These GWs carry information about the interior composition of the NS, that can be used in constraining EoS.  The direct detection of GWs from the binary NS merger event GW170817 by the LIGO-Virgo Collaboration along with its electromagnetic counterparts opened up a new window to multimessenger astronomy~\cite{Abbott_2019_LVK}.
GWs may also be produced by unstable NS oscillation modes such as fluid $f-,p-,g-$ modes or rotational $r$-modes through the Chandrasekhar-Friedman-Schutz (CFS) mechanism. Although GWs from such modes have not yet been detected, even a non-detection imposes important constraints on NS matter properties~\cite{LVK_2022_AllSkyCW,LVK_2022_NarrowCW,LVK_2022_KnownPSR_CW}. 

Matter probed in terrestrial experiments (nuclear and heavy-ion experiments) is isospin symmetric, in contrast with NS matter which is highly isospin asymmetric, and therefore they provide complementary information on the nuclear EoS. As the behaviour of the EoS at large density and isospin asymmetry is unknown, one needs to resort to theoretical EoS models to extrapolate the known information to both higher densities and finite asymmetry. However such an extrapolation is the source of uncertainties in the EoS models. The difference between the energy of pure neutron matter (PNM) and symmetric nuclear matter (SNM), the so-called {\it symmetry energy} and its density dependence are known to show strong correlations with the neutron skin in neutron-rich nuclei as well as with the NS radius~\cite{Reinhard_2016, Chen_2014, piekarewicz_2013, Agrawal_2010_PRC}. 

Many theoretical models have been proposed in the literature to describe the microscopic EoS of NSs, such as the {\it ab initio} and the {\it phenomenological} models~\cite{oertel2017_Rev_ModPhys,Dutra2012,Dutra2014}. In ab-initio models, the many body problem is solved starting with the nuclear properties in vacuum. Chiral Effective Field Theory ($\chi$EFT) is one such microscopic ab-initio framework based on the chiral perturbation theory, that provides a reliable description of pure neutron matter (PNM) at low density (up to 1.4$n_{sat}$)~\cite{drischler2016_PRC}. Recently, these calculations have also been extended to asymmetric nuclear matter (ANM) relevant for application in neutron stars~\cite{Keller_Hebeler_2023prl,Rutherford_2024}. On the other hand, in phenomenological models, the density functional theory is applied to describe the EoS. E.g. in the {\it Relativistic Mean Field} (RMF) model, nucleons participate in strong interaction via meson exchange, and a mean-field approximation is applied for meson fields. In the RMF model, the model parameters are fitted to reproduce nuclear empirical observables at nuclear saturation density.

There also exist various parametric EoS models which model the EoS using functional forms without depending on the underlying nuclear physics, such as piecewise polytropes \cite{Annala_2018,Hebeler2013,Read2009,Gamba2019}, spectral representation~\cite{Fasano2019,Lindblom2018} or the speed-of-sound parametrization~\cite{Tews_2018,Greif2019,Landry2020}. These have been widely applied in numerical relativity simulations as well as for parameter estimation from gravitational waveforms. The nuclear metamodelling technique~\cite{JM1,JM2} employs an expansion of the energy in terms of the nuclear empirical observables, and is by definition model-agnostic, but this parametrization is only valid close to the saturation density.  Recently, non-parametric inference schemes have also been constructed~\cite{Landry2019,Legred2021}, using Gaussian processes trained on nuclear EoSs and extended to include $\chi$EFT calculations~\cite{Essick2020}. The above schemes satisfy general physical constraints such as causality, but the possible degrees of freedom appearing at intermediate or high densities are not included, and the role of the underlying nuclear saturation parameters in controlling the behaviour of NS global observables is not evident. 
A few EoSs have also attempted to combine nuclear + piecewise polytrope parametrizations~\cite{Biswas2021} or supplemented non-parametric priors with nuclear information at low densities to obtain posterior distributions of empirical parameters but the correlations between the nuclear empirical parameters and NS astrophysical observables were not clearly established. 
Parametric EoSs have been used to impose constraints on the EoS by using NS multi-messenger observations within a statistical Bayesian formalism~\cite{pang2021,Coughlin2019,Biswas2021,biswas2021prex,biswas2021bayesian,Dietrich2020,O_Boyle_2020} by matching the low density EoS constrained by theoretical and experimental nuclear physics with parametrized high density EoSs satisfying multi-messenger observational data ~\cite{Capano2020,Tews_2018,Tews_2019a,Tews_2019b,Gandolfi_2019}. 
Some of these works explored correlations among empirical nuclear parameters and the NS structure properties,~\cite{Zhang2019,Xie2019,Carson2019,zimmerman2020measuring,Guven2020} but none of these works performed a study of neutron star oscillations or established a connection between nuclear parameters with the characteristics of oscillation modes.

In a series of recent works~\cite{Ghosh2022a,Ghosh2022b}, we developed a formalism to restrict the parameter space of the nuclear model by implementing state-of-the-art constraints from multidisciplinary physics at different densities, and systematically investigated the role of empirical nuclear parameters on NS observables. This Bayesian-inspired scheme involves varying nuclear parameters within the RMF formalism within their allowed uncertainties, and then imposing constraints from nuclear, heavy-ion experiments and multi-messenger astrophysical observables. 
The correlation studies in these works indicated that the Dirac effective nucleon mass $m^*_D/m$ is the most sensitive parameter that strongly correlates with the NS properties. By performing systematic correlation studies of $f-$mode oscillations within the RMF scheme~\cite{Jaiswal2021,Pradhan2021,Pradhan2022} (for Cowling approximation and full GR-formalism, and recently extended to hot NSs) we also demonstrated that among the nuclear empirical parameters, it is the Dirac effective nucleon mass $m^*_D/m$ which shows the strongest correlation with $f-$mode characteristics. The inverse problem was also demonstrated, i.e. how future observations of GWs from $f-$modes in isolated or binary systems can be used to constrain the nuclear empirical parameters within the RMF scheme~\cite{Pradhan_2023,Pradhan_2024}. \\

\textbf{Aim of this work:} The question naturally arises whether such a correlation of the NS observables with the effective nucleon mass or with other nuclear empirical parameters is physical or arise due to the functional form of the  chosen nuclear (RMF) model. It has been discussed that correlations among empirical parameters found within a specific functional family (such as RMF) may be modified in other functional forms~\cite{Ducoin_2011,Khan_2013,Nazarewicz_2014}. A possible solution was proposed by introducing the meta-modelling (MM) approach~\cite{JM1,JM2} as a representation of the class of continuous EoS models. The meta-model consists of a functional flexible enough to reproduce different functionals obtained from most existing models (non-relativistic or relativistic DFT or ab-initio) within its parameter space. The prior distribution of the meta-model parameters is obtained from a compilation of the different models belonging to a given class.
In order to test the model-dependence, we use the same Bayesian-inspired scheme, and obtain posterior distributions by filtering the meta-model parameter space with the constraint of reproducing experimental observables. The resulting EoS is then an average over the different models, and the artificial parameter correlations arising from the choice of a given model are naturally suppressed.
We then compare our results for the $f-$mode characteristics for the two models within the relativistic Cowling approximation.  In our previous work~\cite{Pradhan2022}, we demonstrated that although the relativistic Cowling approximation overestimates the $f-$mode frequencies by $\sim 20\%$, it captures the correct qualitative features of the $f-$mode properties, such as the correlations with nuclear parameters. 

This paper is organized as follows: in Section~\ref{sec:formalism}, we describe the construction of NS models, from the microscopic description of the EoSs within two different schemes to calculating the global NS observables. The calculation of the $f-$mode oscillations and the scheme for calculation of correlations are also discussed. In Section~\ref{sec:prelim}, preliminary studies of the sensitivity of NS observables to nuclear parameters are performed. The results of the correlation study for both the EoS models is discussed in Section~\ref{sec:results}. Finally, in Section~\ref{sec:discussions}, we discuss the implications of our work and compare our findings with other recent works in the literature.

\section{Formalism}
\label{sec:formalism}

In this section, we describe the detailed formalism adopted for the microscopic as well as macroscopic description of the NS. In Section~\ref{sec:nuclear_parameters}, we define the nuclear empirical parameters. In Section~\ref{sec:eos}, we describe the EoS of the NS core using two different models: the Relativistic Mean Field (RMF) and the Meta-model (MM). In Section~\ref{sec:structure}, we describe the calculation of the global structure properties of NSs. In the Section~\ref{sec:oscillation_modes}, we detail the calculation of $f-$mode characteristics within the relativistic Cowling approximation. Finally, we describe the constraints imposed within the Bayesian-inspired scheme in Section~\ref{sec:Imposing_constraints}. \\

\subsection{Nuclear empirical parameters}
\label{sec:nuclear_parameters}
The extrapolation from nuclear saturation density to higher densities and from symmetric (SNM) to asymmetric nuclear matter (ANM) is introduced in theoretical models. The energy per nucleon can be expanded around the isospin asymmetry coefficient $\delta = 0$,
\begin{equation}
\label{eq:ener_expansion}
    e(x, \delta) = e_{is}(x) + \delta^2e_{iv}(x)
\end{equation}
where $x=(n - n_{sat})/3n_{sat}$ is the dimensionless density, and $\delta = (n_1/n) = (n_n - n_p)/(n_n + n_p)$ is the isospin asymmetry parameter. $n,\; n_n,\;n_p$ are the number densities of baryon, neutron, and proton, respectively. Nuclear saturation density ($n_{sat}$) is the density at which the energy per nucleon is minimum for symmetric nuclear matter. The isoscalar and isovector contributions to the energy per nucleon,  $e_{is}(x)$ and $e_{iv}(x)$ respectively, can be expanded around saturation density ($x = 0$) as
\begin{center}
\resizebox{0.96\hsize}{!}{$
\begin{aligned}
e_{is}(x) = &E_{sat} + \frac{1}{2!}K_{sat}x^2 + \frac{1}{3!}Q_{sat}x^3 + \frac{1}{4!}Z_{sat}x^4 +...~,\\
e_{iv}(x)  = &J_{sym} + L_{sym} x + \frac{1}{2!}K_{sym} x^2 + \frac{1}{3!}Q_{sym} x^3 \nonumber\\
&+ \frac{1}{4!}Z_{sym}x^4 +...~,
\end{aligned}
$}
\end{center}
\vspace{-2em}
\begin{equation}
\tag{\theequation}\label{eq:isoscalar_isovector_ener}
\stepcounter{equation}
\end{equation}

where $n_{sat}$, $E_{sat}$, $K_{sat}$, $J_{sym}$, $L_{sym}$, $K_{sym}$, $Q_{sat}$, $Q_{sym}$, $Z_{sat}$, $Z_{sym}$ and $m^*/m$ are the empirical nuclear parameters measured in terrestrial experiments. $E_{sat}, \; K_{sat}, \; J_{sym}$, $L_{sym}$, and $K_{sym}$ are the energy per nucleon (or binding energy per nucleon), the incompressibility modulus, the symmetry energy, the slope of the symmetry energy, and the isovector incompressibility respectively, all calculated at saturation density. The parameter $m^*/m$ is the effective nucleon mass at $n_{sat}$. The higher-order parameters $Q_{sat}$, $Q_{sym}$, $Z_{sat}$, and $Z_{sym}$ correspond to the isoscalar skewness, isovector skewness, isoscalar kurtosis, and isovector kurtosis respectively. The empirical parameters at saturation density are extracted from measurements of the neutron skin thickness of $^{208}\text{Pb}$ and $^{48}\text{Ca}$, electric dipole polarizability $\alpha_D$, isoscalar giant monopole resonances, nuclear masses, and several other experimental data~\cite{Shlomo_2006,Centelles2009,Warda2009,Roca2011}.

\subsection{Microscopic description (EoS)} 
\label{sec:eos}

\subsubsection{RMF Model}
The Relativistic Mean Field (RMF) model is one of the commonly used phenomenological models to describe the nuclear EoS of neutron stars. In the RMF model, the nucleon-nucleon (NN) interactions are described by the exchange of mesons~\cite{Chen_2014,Hornick2018}. The total effective Lagrangian density in the non-linear RMF model can be written as follows.
\begin{center}
\resizebox{0.96\hsize}{!}{$
\begin{aligned}
    \mathcal{L} &= \sum_N \Bar{\Psi}_N[(i\gamma_{\mu}\partial^\mu - m) + g_\sigma \sigma - g_\omega \gamma^\mu\omega_\mu - \frac{g_\rho}{2}\gamma^\mu \tau \cdot \rho_\mu]\Psi_N \nonumber \\ 
    &+ \frac{1}{2}(\partial_\mu\sigma\partial^\mu\sigma - m_\sigma^2\sigma^2) - \frac{1}{3}bm(g_\sigma\sigma)^3 - \frac{1}{4}c(g_\sigma\sigma)^4 \nonumber\\
    &- \frac{1}{4}\omega_{\mu\nu}\omega^{\mu\nu} + \frac{1}{2}m_\omega^2\omega_\mu\omega^\mu + \frac{\zeta}{4!}(g_\omega^2\omega_\mu \omega^\mu)^2 \nonumber\\
    &-\frac{1}{4}\rho_{\mu\nu}\cdot\rho^{\mu\nu} + \frac{1}{2}m_\rho^2\rho_\mu\rho^\mu  + \Lambda_\omega (g_\rho^2\rho_\mu \cdot \rho^\mu)(g_\omega^2 \omega_\mu\omega^\mu) 
\end{aligned}
$}
\end{center}
\vspace{-2em}
\begin{equation}
\tag{\theequation}\label{eq:lagrangian}
\stepcounter{equation}
\end{equation}

where $\Psi_N$ is the Dirac field for nucleons $N$, $\gamma^\mu$ and $\vec{\tau}$ denote the Dirac and Pauli matrices. $\sigma, \omega_\mu, \vec{\rho}_\mu$ are scalar, vector and isovector mesonic fields respectively. Here $m$, $m_\sigma$, $m_\omega$ and $m_\rho$ are the average nucleon mass and masses of the $\sigma$, $\omega$ and $\rho$ mesons respectively.\\

The isoscalar coupling constants $g_\sigma$ and $g_\omega$ of $\sigma-N$ and $\omega-N$ interactions fix the energy per particle and ground state density. The isovector coupling strength $g_\rho$  and $\Lambda_\omega$ of $\rho-N$ and $\omega-\rho$ interactions control the density-dependent nuclear symmetry energy. The $\sigma$ meson self-interaction strengths $b$ and $c$ are important for reproducing the correct incompressibility of nuclear matter at saturation. In our study, we set the quartic $\omega$-coupling constant $\zeta$ as zero as it is known to soften the EoS, which conflicts with the $2M_\odot$ constraints from pulsar data.\\

Using the total effective Lagrangian density from equation~(\ref{eq:lagrangian}) in the Euler-Lagrange equations, one can obtain the energy density
\begin{center}
\resizebox{0.96\hsize}{!}{$
\begin{aligned}
     \epsilon = &\sum_N \frac{1}{8\pi^2}\left[k_{F_N}E_{F_N}^3 + k_{F_N}^3E_{F_N} - m^{*4}_D\;ln\left(\frac{k_{F_N} + E_{F_N}}{m^*_D}\right)\right ] \nonumber \\
     &+ \frac{1}{2}m_\sigma^2\bar{\sigma}^2  + \frac{1}{3}bm(g_\sigma \bar{\sigma})^3 + \frac{1}{4}c (g_\sigma \bar{\sigma})^4 + \frac{\zeta}{8}(g_\omega \bar{\omega})^4\nonumber\\
     &+ \frac{1}{2}m_\omega^2\bar{\omega}^2 + \frac{1}{2}m_\rho^2\bar{\rho}^2 
      + 3\Lambda_\omega (g_\rho g_\omega\bar{\rho}\bar{\omega})^2
\end{aligned}
$}
\end{center}
\vspace{-2em}
\begin{equation}
\tag{\theequation}\label{}
\stepcounter{equation}
\end{equation}
where the Dirac effective mass of the nucleon $m^*_D$ is defined as
\begin{equation}
\label{eq:eff_mass}
    m^*_D = m - g_\sigma \bar{\sigma}~.
\end{equation}
$\bar{\sigma}$, $\bar{\omega}$, and $\bar{\rho}$ denote the expectation values of the meson fields in the mean field approximation. Here $k_{F_N}$ is the Fermi momentum of nucleon and Fermi energy is given by $E_{F_N} = \sqrt{k_{F_N}^2 + m^{*2}_D}$.
 The pressure can be obtained from the Gibbs-Duhem relation 
 \begin{equation}
     P = \sum_N \mu_N n_N - \epsilon~,
 \end{equation}
 where chemical potential $\mu_N$ is defined as \\
 \begin{equation}
     \mu_N = E_{F_N} + g_\omega \bar{\omega} + \frac{g_\rho}{2} \tau_{3N} \bar{\rho}~.
 \end{equation}

 \subsubsection{Meta Model (MM)}
The {\it Meta-model} is a non-relativistic phenomenological model to study the EoS of nuclear matter inside neutron stars \cite{JM1, JM2}. In this metamodelling approach, the energy per particle is defined as
 \begin{equation}
     e^N(n, n_1) = t^{FG*}(n, n_1) + V^N(n, n_1)~,
 \end{equation}
 Here, the kinetic energy term $t^{FG*}(n, n_1)$ and the potential energy term $V^N(n, n_1)$ (up to order N of the Taylor series) are defined as follows:
 \begin{align}
    t^{FG*}(n, n_1) = &\frac{t^{FG}_{sat}}{2}\left (\frac{n}{n_{sat}}\right )^{2/3}\left[\left (1 +  \kappa_{sat}\frac{n}{n_{sat}}\right)f_1(\delta) \right. \nonumber  \\
    &\left. + \kappa_{sym} \frac{n}{n_{sat}}f_2(\delta) \right ]~,\\
    V^N(n, n_1) = &\sum_{\alpha \ge 0}^N\frac{1}{\alpha !}(v_\alpha^{is} + v_\alpha^{iv}\delta^2)x^\alpha u_\alpha^N(x)~,
 \end{align}
 where $t^{FG}_{sat} = (3\hbar^2/(10 m))(3\pi^2/2)^{2/3} n_{sat}^{2/3}$ is the kinetic energy per nucleons for symmetric matter at saturation, $m$ is the average nucleon mass. 
 The asymmetry coefficient $\delta = [0,1]$ and the dimensionless density $x$ are as defined earlier in Section~\ref{sec:nuclear_parameters}. $v_\alpha^{is}$ and $v_\alpha^{iv}$ are the isoscalar and isovector model parameters, respectively. Here, $$u_\alpha^N(x) = 1 - (-3x)^{N+1-\alpha}\;exp\;(-bn/n_{sat})$$ is the correction factor introduced to reproduce the correct behavior of energy density at zero density (model ELFc) \cite{JM1}. The parameter $b = 10 \ln (2) \approx 6.93$. The functions $f_1$ and $f_2$ are defined as follows,
 \begin{align}
     f_1(\delta) = &(1 + \delta )^{5/3} + (1 - \delta)^{5/3}\\
     f_2(\delta) = & \delta [(1 + \delta )^{5/3} - (1 - \delta)^{5/3}]~.
 \end{align}
 The Landau effective nucleon mass ($m^*_{L,N}$) in MM is defined as 
 \begin{equation}
     \frac{m}{m^*_{L,N}} = 1 + (\kappa_{sat} + \tau_{3N} \kappa_{sym} \delta) \frac{n}{n_{sat}}~.
 \end{equation}

 Here $\tau_3 = +1$ and $-1$ for neutrons and protons, respectively. It is important to note that the definition of the effective mass (Landau effective mass) in the meta-model is different from the RMF model (Dirac effective mass)~\cite{Raduta2021}. The parameters $\kappa_{sat}$ and $\kappa_{sym}$ are functions of Landau effective mass at saturation for SNM ($m^*_L/m$), and isospin mass splitting at saturation ($\Delta m^*_L/m$) for pure neutron matter (PNM)~\cite{Alvarez2020a, Alvarez2020b}. These can be written as follows,
 \begin{align}
     \kappa_{sat} =& \frac{m}{m^*_L} - 1~,  \nonumber\\
     \kappa_{sym} = & \frac{1 - \sqrt{1 + \left( \frac{\Delta m^*_L}{m^*_L} \right)^2 }}{\frac{\Delta m^*_L}{m}}~.
 \end{align}

 For small values of isospin splitting ($\Delta m^*_L/m$), $\kappa_{sym} \to -\frac{1}{2}\frac{\Delta m^*_L}{m}\left( \frac{m}{m^*_L}\right)^2$.\\
 
 Now, one can derive the relations between the model parameters and the nuclear empirical parameters at saturation. For isoscalar model parameters,
 \begin{align}
   v^{is}_0 =& E_{sat} - t^{FG}_{sat}(1 + \kappa_{sat}) \nonumber \\
   v_1^{is} = & -t^{FG}_{sat} (2 + 5 \kappa_{sat}) \nonumber \\
   v_2^{is} =& K_{sat} - 2t^{FG}_{sat}(-1+5\kappa_{sat}) \nonumber \\
   v_3^{is} =& Q_{sat} - 2t^{FG}_{sat}(4-5\kappa_{sat}) \nonumber \\
   v_4^{is} =& Z_{sat} - 8t^{FG}_{sat}(-7+5\kappa_{sat})~.
 \end{align}
 Similarly, for isovector parameters,
 \begin{align}
     v_0^{iv} =&J_{sym} - \frac{5}{9}t^{FG}_{sat}[1 + (\kappa_{sat}+3\kappa_{sym})] \nonumber \\
   v_1^{iv} = & L_{sym} - \frac{5}{9}t^{FG}_{sat} [2 + 5 (\kappa_{sat}+3\kappa_{sym})]\nonumber \\
   v_2^{iv} =& K_{sym} - \frac{10}{9}t^{FG}_{sat}[-1+5(\kappa_{sat}+3\kappa_{sym})] \nonumber \\
   v_3^{iv} =& Q_{sym} - \frac{10}{9}t^{FG}_{sat}[4-5(\kappa_{sat}+3\kappa_{sym})]  \nonumber \\
   v_4^{iv} =& Z_{sym} - \frac{40}{9}t^{FG}_{sat}[-7+5(\kappa_{sat}+3\kappa_{sym})]~.
 \end{align}
 
The pressure is the first derivative of the energy per nucleon, which is denoted as $P = n^2\left [\frac{\partial e(n, n_1)}{\partial n}\right ]$.

\subsection{Macroscopic properties}
\label{sec:structure}

\subsubsection{Mass and Radius of NS}
\label{sec:MR_calc}

Once the microscopic EoS is obtained using the RMF or Meta model, the mass and radius of a spherically symmetric non-rotating NS can be obtained by solving Tolman-Oppenheimer-Volkoff (TOV) equations~\cite{Glendenning_book,Schaffner-Bielich_book}.
\begin{align}
    &\frac{dm(r)}{dr} = 4\pi\epsilon(r) r^2 \nonumber\\
    &\frac{dp(r)}{dr} = - \frac{[p(r) + \epsilon(r)][m(r) + 4\pi r^3 p(r)]}{r(r - 2m(r))}~.
\end{align}
Here, the natural units (c = $\hbar $ = G = 1) have been used. To solve these equations, the boundary conditions are $m(r=0) = 0,\; m(r=R) = M, \;p(r=0) = p_c$, and $p(r = R) = 0$, where $M$ and $R$ are the total mass and radius of the neutron star corresponding to the central pressure $p_c$, respectively. \\

\subsubsection{Tidal deformability}
\label{sec:tidal_deformability}
In a binary NS system, one compact object gets distorted because of the tidal field of the other. The strength of distortion is referred to as the {\it tidal deformability} $(\lambda)$, and it measures the induced quadrupole moment $Q_{ij}$ due to the tidal field $\epsilon_{ij}$ of the companion star,
\begin{equation}
    Q_{ij} = -\lambda \epsilon_{ij}~.
\end{equation}
The tidal deformability $\lambda$ is connected to the dimensionless $l = 2$ tidal love number $k_2$ by the relation,
\begin{equation}
    \lambda = \frac{2}{3}k_2 R^5~.
\end{equation}
The dimensionless tidal deformability $\Lambda$ is defined as
\begin{equation}
\label{eq:lamda_eq}
    \Lambda = \frac{2k_2}{3 C^5} = \frac{2k_2}{3}\left (\frac{R}{M} \right )^5
\end{equation}
where $C = M/R$ is the compactness parameter of the NS. In order to calculate the tidal deformability, one has to solve a set of differential equations coupled with the TOV equation as done in~\cite{Damour2009, Hinderer2008_PRD}.

\subsection{Calculation of oscillation modes}
\label{sec:oscillation_modes}

Non-radial oscillation modes in stars have been studied in great detail~\cite{Andersson_1998,Benhar_2004}. In the non-relativistic framework, Cowling first outlined the procedure to analyze these modes~\cite{Cowling_1941}, while the general relativistic (GR) treatment was developed by Thorne and Campollataro~\cite{Thorne_1967}. In full GR, one must account for the metric perturbations while solving the perturbed fluid equations, which are time-consuming and computationally expensive. However, within the Cowling approximation, which is valid in weak gravitational fields, these metric perturbations can be neglected.
 Many works in the literature~\cite{Doneva2013,Sandoval2018,Flores2017,Pradhan2021}, have employed this simplification for studies of $f-$modes. It has been demonstrated that the oscillation frequency for the $f-$mode obtained using the Cowling approximation is real and overestimated by less than $20\%$ from those calculated using the full GR treatment~\cite{Yoshida1997,Pradhan2022}. In the Cowling approximation a calculation of the damping time of QNMs is not possible.
\\

In the present work, we focus on the Cowling approximation, and hence the metric for the spherically symmetric background is given by,
\begin{equation}
    ds^2 = -e^{2\Phi(r)} dt^2 + e^{2\Lambda(r)}dr^2 + r^2d\theta^2 + r^2sin^2\theta d\phi^2
\end{equation}

The equation governing the fluid oscillations with perturbations can be derived by varying the conservation of the energy-momentum tensor. The fluid's Lagrangian displacement vector is given by
\begin{equation}
    \zeta^i = \left(e^{-\Lambda(r)}W, -V\partial_\theta, -Vsin^{-2}\theta \partial_\phi \right)r^{-2} Y_{lm}(\theta, \phi)~,
\end{equation}

where $Y_{lm}$ is the spherical harmonics. Assuming the harmonic time dependency, two functions $W(r)$ and $V(r)$ satisfy the following coupled equations~\cite{Sotani_2011,Pradhan2021}:
\begin{align}
    \frac{dW(r)}{dr} = \,&\frac{d\epsilon}{dp} \left[ \omega^2 r^2 e^{\Lambda (r) - 2\phi(r)}V(r) + \frac{d\Phi(r)}{dr} W(r) \right] \nonumber \\
    & -l(l+1)e^{\Lambda (r)} V(r)~,\\
    \frac{dV(r)}{dr} =\; & 2\frac{d\Phi(r)}{dr} V(r) - \frac{1}{r^2} e^{\Lambda (r)} W(r)~,
\end{align}
where
\begin{equation}
    \frac{d\Phi(r)}{dr} = -\left[\frac{1}{\epsilon(r) + p(r)}\right]\frac{dp}{dr}~.
\end{equation}
The functions $W(r)$ and $V(r)$ at the centre of the NS behave as
\begin{equation}
    W(r) = A r^{l+1};\; V(r) = -\frac{A}{l}r^l~,
\end{equation}
where $A$ is an arbitrary constant. Given the EoS $p=p(\epsilon)$, one can determine the eigen frequencies ($\omega$) satisfying the following boundary condition 
 at the surface for each $l$ value,
\begin{equation}
    \omega^2e^{\Lambda(R) - 2\Phi(R)}V(R) + \frac{1}{R^2}\frac{d\Phi(r)}{dr}\Big|_{r=R} W(R) = 0~,
\end{equation}

where $R$ is the radius of the Neutron Star.

\subsection{Imposing constraints on the EoS}
\label{sec:Imposing_constraints}

In order to study the correlations between NS properties and nuclear empirical parameters, in this work we apply the cut-off filter scheme inspired by a Bayesian formalism following our recent works~\cite{Ghosh2022a,Ghosh2022b}. In this scheme, the nuclear parameters are varied randomly within their allowed uncertainty range compatible with nuclear experimental data, to generate a flat EoS prior set. Then `cut-off' filters are applied i.e. physical constraints are imposed at different densities and the EoS posteriors are generated. Then correlation studies are performed using these posteriors. In this study, we use two cut-off filters described in the following:\\

\subsubsection{$\chi$EFT at low density}
\label{sec:CEFT_constraint}

In the baryon density range of $0.08 - 0.21\;fm^{-3}$, the EoS for pure neutron matter (PNM) can be calculated using the Chiral Effective Field Theory ($\chi$EFT)~\cite{drischler2016_PRC} calculations. This theory is based on the microscopic calculation of many-body interactions among nucleons considering order-by-order expansions. We use this constraint on the pure neutron matter (PNM) EoS priors at lower densities. The required EoS parameter sets should comply with the results of $\chi$EFT within the mentioned baryon density region.\\

\begin{figure}[ht]
    \centering
    \includegraphics[width=1\linewidth]{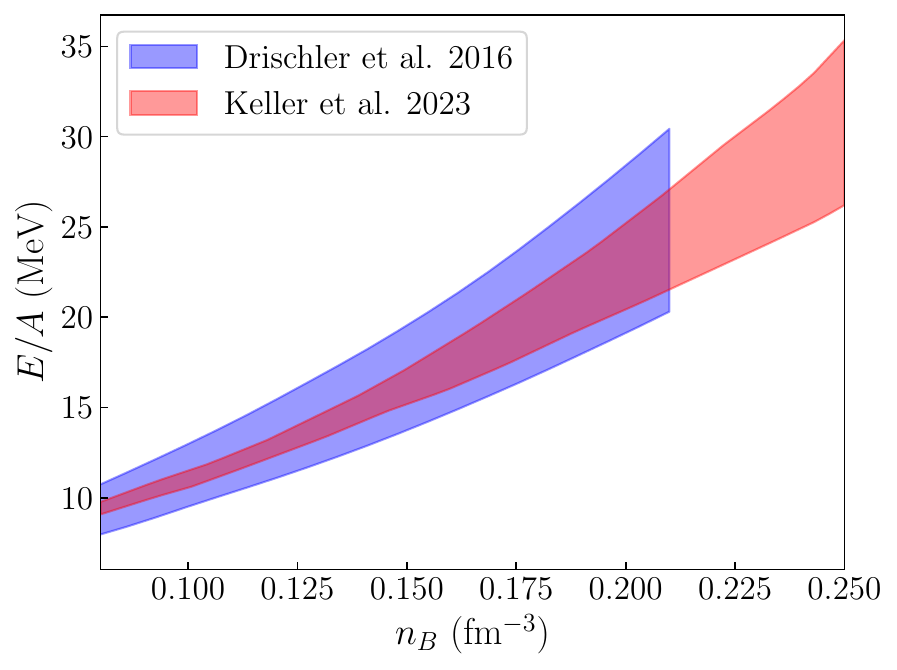}
    \caption{The variation of energy per baryon ($E/A$) with baryon number density ($n_B$) for pure neutron matter has been plotted. The bands display theoretical $\chi$EFT uncertainty at next-to-next-to-next-to-leading order (N$^3$LO). These bands are taken from~\cite{drischler2016_PRC, Keller_Hebeler_2023prl}. See text for further details.}
    \label{fig:xEFT_pnm_band_comparison}
\end{figure}

Recently, $\chi$EFT calculations have been extended up to $0.25\;fm^{-3}$ (see Figure~\ref{fig:xEFT_pnm_band_comparison}) for arbitrary proton fractions and temperatures, and using nonparametric EoS
calculations to $\beta$-equilibrated neutron star matter~\cite{Keller_Hebeler_2023prl,Keller_Wellenhofer_2021prc}.
In this work, in order to compare with previous results, we first employ the $\chi$EFT results of Drischler~et~al.~(2016)~\cite{drischler2016_PRC} for PNM. We then impose the new constraints from Keller~et~al.~(2023)~\cite{Keller_Hebeler_2023prl} for PNM and comment on how they affect the correlations.

\subsubsection{Astrophysical Observations at high density}
\label{sec:Astro_constraint}
Multi-messenger (electromagnetic and gravitational wave) observations provide information about the EoS at high density. Some of the crucial state-of-the-art multi-messenger constraints on the NS EoS are the following:\\

i) Pulse profile modelling (PPM) of X-ray data by NICER allows precise measurements of the highest observed NS mass as well as its radius. Until recently, mass-radius inferences were published for the millisecond pulsars PSR J0030+0451~\cite{Miller_2019,Riley_2019} and PSR J0740+6620~\cite{Miller_2021,Riley_2021}.
These observations suggested that the pulsar with the highest maximum observed mass is PSR J0740+6620, with a mass of $2.08^{+0.07}_{-0.07} \; M_\odot$~\cite{Fonseca_2021}. This imposes an important constraint on the maximum mass predicted by theoretical EoS models, which should be able to explain the highest observed NS mass. \\

Very recently, NICER published inference of the mass and radius for PSR J0437-4715~\cite{Choudhury_2024} as well as reanalysis of previous results, using upgraded PPM techniques and instrument models, providing tighter mass-radius constraints~\cite{Salmi_2024, Vinciguerra_2024, Rutherford_2024}. We use the lower limit of $2.01 M_{\odot}$ for the maximum
NS mass~\cite{Fonseca_2021}. We also checked the compatibility of the mass-radius posteriors with the old and new NICER constraints. 
\\

ii) Another important constraint comes from the binary merger event GW170817 \cite{Abbott_2019_LVK}. It limits dimensionless tidal deformability ($\Lambda$) for 1.4$M_\odot$ NS such as $\Lambda < 720$. The correlation (Eq.~\ref{eq:lamda_eq}) between $\Lambda$ with radius ($R$) allows to put important constraints on its value, such that $R < 13.5$ km. 
\\

It is important to note here that there exists an apparent tension between GW data and the astrophysical data from NICER. For the constraint solely from the observation of massive pulsars (PSR J0740+6620), the preferred EoS has to be stiff irrespective of the EoS parameterization.  Adding the constraint only from GW events makes the preferred EoS softer, while further demanding compatibility with NICER results makes the EoS stiffer again, although lesser than that from only pulsar mass observation.
In~\cite{Guven2020}, an apparent tension was found between the empirical parameter based meta-modelling of neutron stars and astrophysical observations. In their article,~\cite{Biswas2021} showed that this tension is an artifact that arises on applying empirically parameterized EoS to high densities relevant for astrophysical observations, which shows preference for a different characterization at high densities.\\

We also mention that in this work do not impose the constraints from heavy-ion data in this work as they are known to be model dependent (see discussions in~\cite{Ghosh2022a}).

\subsubsection{Correlations}
\label{sec:Pearson_corr}
In our study, we take the variables to be random within the allowed uncertainty range to generate a uniformly distributed prior set. Using the posteriors, we study the physical correlations between the nuclear parameters as well as the NS observables such as $R_{1.4M_{\odot}}$, $\Lambda_{1.4M_{\odot}}$, $f_{1.4M_\odot}$, $R_{2.0M_{\odot}}$, $\Lambda_{2.0M_{\odot}}$, and $f_{2.0M_\odot}$ i.e. radius, tidal deformability, and $f-$mode frequency of 1.4 and 2.0$M_{\odot}$ NSs, respectively. Following~\cite{Ghosh2022a,Ghosh2022b}, we use Pearson's linear correlation coefficient, defined as follows,\\
\begin{equation}
    R_{XY} = \frac{Cov(X, Y)}{S(X)S(Y)}
\end{equation}
where $Cov(X, Y)$ is the covariance of the two random variables $X$ and $Y$. The $S(X)$ and $S(Y)$ are the standard deviations of $X$ and $Y$, respectively. 
\\

In our previous work~\cite{Ghosh2022a}, the correlations applying filters from $\chi$EFT and NS astrophysical observables were studied both with and without statistical weighting. It was shown that the inclusion of weights does not qualitatively affect the correlations. Further, experimental data may be model dependent or have large uncertainties and the inclusion of incorrect statistical weights may introduce a false confidence in the results. Therefore in this work, we apply hard cut-off filter constraints and do not include statistical weights, following other works~\cite{Ghosh2022b}.

\section{Preliminary studies}
\label{sec:prelim}

To investigate the role of the nuclear empirical parameters on NS observables, as a preliminary check we first perform a sensitivity study where the parameters are varied individually within the uncertainty range allowed by current experimental data. For comparison, we employ the two different models, i.e., RMF and Meta-model, and reproduce the results from previous studies. The results are discussed in detail below.

\subsection{Sensitivity study using RMF}

In order to test the numerical scheme, we first perform a sensitivity study using the RMF parameter set given in Table \ref{tab:RMF_params_uncertainties} considered in our previous work~\cite{Ghosh2022a,Hornick2018}. Using the RMF set, we examine the behaviour of the EoS as well as NS observables such as mass, radius, and tidal deformability, $f-$mode frequency to the variation of the nuclear empirical parameters within the uncertainty region. We use the relativistic `HS-DD2'~\cite{HEMPEL_2010} EoS for the crust to determine the NS macroscopic properties in sensitivity and correlation studies within the RMF framework. It is known that a non-unified EoS model may introduce errors in the estimation of the NS radius~\cite{Fortin_2017}. However, we verified the maximum error in radius estimation using the non-unified EoS model. We obtained thermodynamically consistent unified crust for the $\beta$-equilibrated core EoS using the \texttt{CUTER} code~\cite{Davis_2024}. From this comparison, we observed that the use of non-unified crust affects the radius for the low mass NSs, but the relative error in radius for the canonical mass NS (1.4M$_{\odot}$) is less than 1\%. We also obtained negligible effects on the dimensionless tidal deformability and $f$-mode frequencies.\\

\begin{table}[ht]
\resizebox{\linewidth}{!}{
\begin{tabular}{  c c c c c c c } 
 \hline \hline
 Model & $n_{sat}$ & $E_{sat}$ & $K_{sat}$ & $J_{sym}$ & $L_{sym}$ & $m^*_D/m$  \\ 
 & $(fm^{-3})$ & (MeV) & (MeV) & (MeV) & (MeV) & \\ 
 [0.5ex]
 \hline
 HTZCS & 0.150 & -16.0 & 240 & [30, 32] & [40, 60] & [0.55, 0.75] \\ [1ex]
 \hline \hline
\end{tabular}
}
\caption{The nuclear empirical parameters and their variations at saturation density for the RMF model. These values are taken from~\cite{Hornick2018}.}
\label{tab:RMF_params_uncertainties}
\end{table}

As the uncertainty in the saturation density $n_{sat}$, energy per particle at saturation $E_{sat}$ and compressibility $K_{sat}$ are smaller compared to the other parameters, these are not varied. We study the individual effects of the other parameters i.e., symmetry energy $J_{sym}$, its slope $L_{sym}$, and the Dirac effective nucleon mass $m^*_D/m$. We find that the NS structure properties (mass, radius, $f-$mode frequency) are highly sensitive to the Dirac effective nucleon mass among all the nuclear parameters, as seen in previous works~\cite{Jaiswal2021,Pradhan2021,Pradhan2022} (figures not provided). \\


\subsection{Sensitivity study using MM}

In this section, we repeat the sensitivity study within the Meta-model framework by varying the nuclear parameters independently within their uncertainties, as given in Table~\ref{tab:MM_params_uncertainties}. The Meta-model is described by a Taylor series expansion in density. So, one can observe that the empirical parameters at different orders play a crucial role at different densities. The higher-order parameters have an impact away from saturation density. As reported in~\cite{JM1}, we find that among the lower order parameters, the effects of $E_{sat}$, $K_{sat}$, and $J_{sym}$ on the energy per particle, pressure, and symmetry energy are negligible for SNM ($\delta = 0)$, PNM ($\delta =1$), and a special case of ANM ($\delta = 0.5$). However the EoS shows significant sensitivity to $n_{sat}$, $L_{sym}$, and higher-order parameters such as $K_{sym}$, $Q_{sat}$, $Z_{sat}$, $Q_{sym}$, $Z_{sym}$. But the EoS is insensitive to variations in Landau effective mass ($m^*_L/m$), and the isospin mass splitting ($\Delta m^*_L/m$).

\subsubsection{Sensitivity study of higher-order parameters in NS observables (MM)}
\label{sec: MM_sensitivity_study}
\begin{figure*}[htbp]
    \centering
    \includegraphics[scale = 0.6]{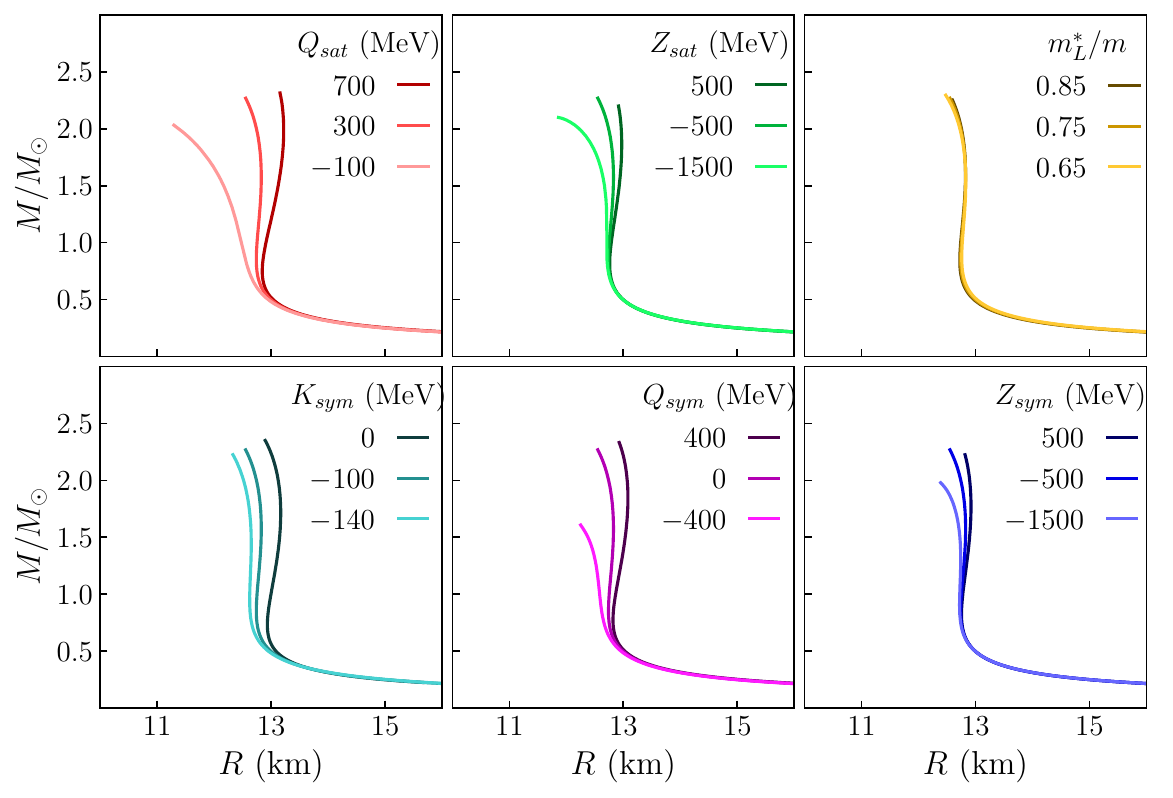}
    \caption{Mass-radius $(M$-$R)$ relations due to variation in higher-order parameters ($K_{sym}$, $Q_{sat}$, $Z_{sat}$, $Q_{sym}$, $Z_{sym}$) and the Landau effective mass ($m^*_L/m$) within Meta-model framework. The individual parameters are varied around their central values (when others are fixed at the central values) within their uncertainties from Table~\ref{tab:MM_params_uncertainties}.}
    \label{fig: MR_var_MM}
\end{figure*}

\begin{figure*}[htbp]
    \centering
    \includegraphics[scale = 0.6]{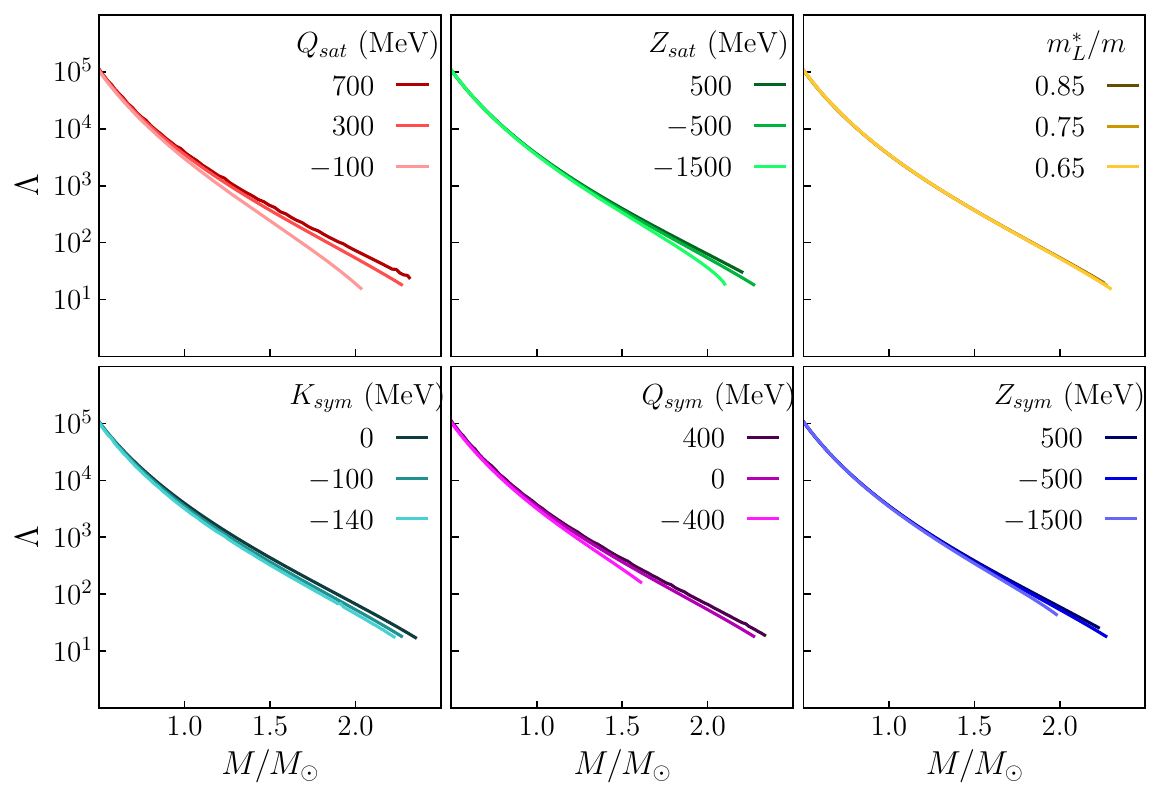}
    \caption{Dimensionless tidal deformability ($\Lambda$) as a function of Neutron Star mass for variation in $K_{sym}$, $Q_{sat}$, $Z_{sat}$, $Q_{sym}$, $Z_{sym}$, and $m^*_L/m$ within Meta-model framework. The individual parameters are varied around their central values (when others are fixed at the central values) within their uncertainties from Table~\ref{tab:MM_params_uncertainties}.}
    \label{fig: tidaldeform_var_MM}
\end{figure*}

\begin{figure*}[htbp]
    \centering
    \includegraphics[scale = 0.6]{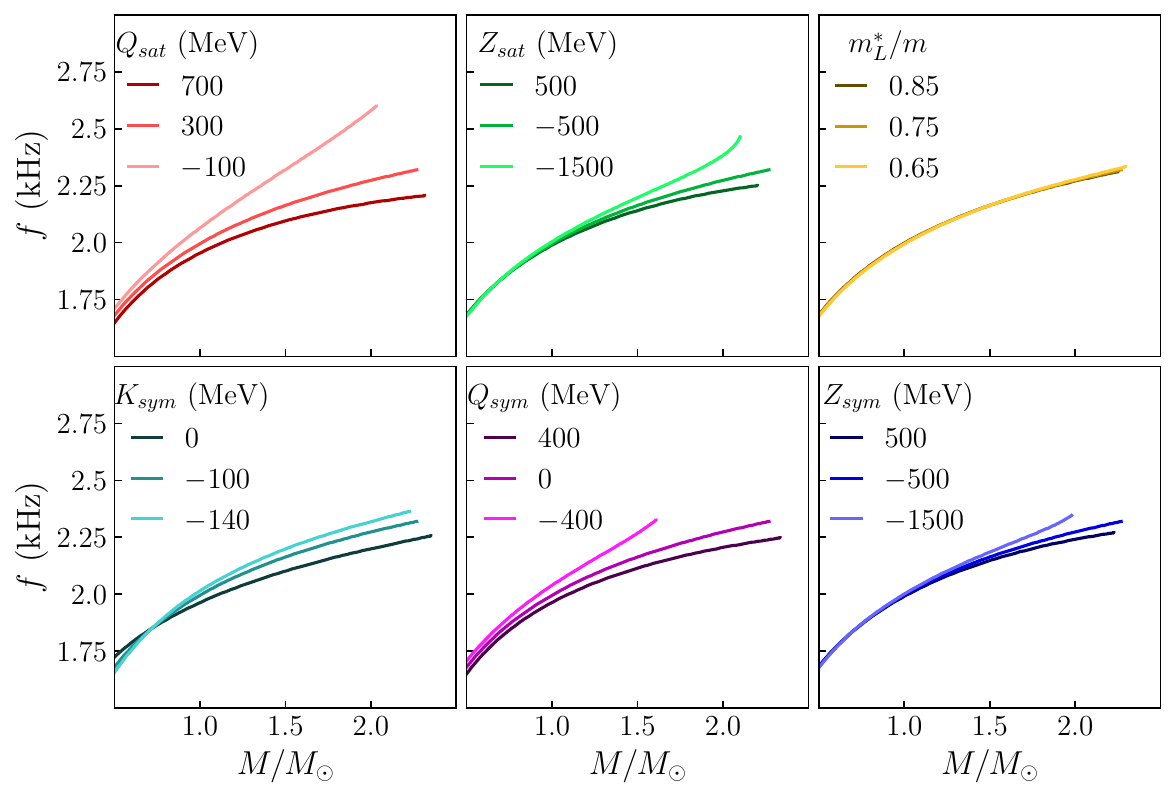}
    \caption{$f-$mode frequency as a function of Neutron Star mass for variation in $K_{sym}$, $Q_{sat}$, $Z_{sat}$, $Q_{sym}$, $Z_{sym}$, and $m^*_L/m$ within Meta-model framework. The individual parameters are varied around their central values (when others are fixed at the central values) within their uncertainties from Table~\ref{tab:MM_params_uncertainties}. }
    \label{fig: fmode_var_MM}
\end{figure*}

Next, we perform a detailed study of the sensitivity of the NS mass, radius, dimensionless tidal deformability and frequency of $f-$mode oscillations, to the higher-order parameters i.e., $K_{sym},Q_{sat},Z_{sat},Q_{sym}, Z_{sym}$ as well as Landau effective mass. For this, we calculate the $\beta$-equilibrated EoS within the MM framework for the $npe\mu$ matter in the NS core. As Meta-model is a non-relativistic model, the core EoS may become acausal at high density. To circumvent this unphysical scenario, we limit the core EoS up to a certain density beyond which the causality principle no longer holds, while also ensuring that the symmetry energy remains positive for all considered densities. Though some non-relativistic crust EoSs such as SLy4 and BPS are available, it is well known that the crust-core matching may result in uncertainties in the NS radii up to 30\% \cite{Fortin_2016,Fortin_2017}. For this work in MM, we use the publicly available $\texttt{CUTER}$ code~\cite{Davis_2024} to construct the NS crust EoS corresponding to the NS core EoS in a thermodynamically consistent way and obtain a unified EoS. Solving the TOV equations described in Section~\ref{sec:structure}, we calculate the mass and radius of the NS. The $f-$mode frequency is obtained using the Cowling approximation, as described in Section~\ref{sec:oscillation_modes}. The effects of variation of higher-order parameters as well as the Landau effective mass on mass-radius, tidal deformability, and $f-$mode frequency are shown in Figures~\ref{fig: MR_var_MM}, \ref{fig: tidaldeform_var_MM}, and \ref{fig: fmode_var_MM} respectively. All higher-order parameters influence NS radius, tidal deformability, and $f-$ mode frequency. However, $Q_{sat}$ has a greater impact on $R_{1.4M_\odot}$ and $R_{2.0M_\odot}$ compared to the other parameters $Z_{sat}$ and $Z_{sym}$, which affect the radius for heavier NSs. The lower limit of $Q_{sym}$ uncertainty is unable to support a  $2M_\odot$ NS. The radius and tidal deformability increase with increasing the value of the higher-order nuclear parameters, whereas the $f-$mode frequency decreases. Thus one can observe that the high-density behaviour in Meta-model is governed by all the higher-order parameters ($K_{sym}$, $Q_{sat}$, $Z_{sat}$, $Q_{sym}$, $Z_{sym}$). An important distinction in the Meta-model is that the uncertainty in the Landau effective nucleon mass does not have an appreciable impact on NS observables, unlike in the RMF model, where NS observables are highly sensitive to the Dirac effective nucleon mass. For the meta-model version of the chiral Hamiltonian H2, a comparable sensitivity study was performed by~\cite{Grams_2024}.

\begin{table*}[ht]
\centering
\resizebox{\linewidth}{!}{
\begin{tabular}{ c c c c c c c c c c c c c} 
 \hline \hline
Model & $n_{sat}$ & $E_{sat}$ & $K_{sat}$ & $Q_{sat}$ & $Z_{sat}$ & $J_{sym}$ & $L_{sym}$ & $K_{sym}$ & $Q_{sym}$ & $Z_{sym}$ & $m^*_L/m$ & $\Delta m^*_L/ m$\\
 & $(fm^{-3})$ & (MeV) & (MeV) & (MeV) & (MeV) & (MeV) & (MeV) & (MeV) & (MeV) & (MeV) & & \\ [0.5ex] 
 \hline 
 MM & 0.155 & -15.8 & 230 & 300 & -500 & 32  & 60 & -100 & 0 & -500 & 0.75 & 0.1\\ [1ex] \hline 
 variation & 0.005 & 0.3 & 20 & 400 & 1000 & 2 & 15 & 100 & 400 & 1000 & 0.1 & 0.1\\  [1ex]
 \hline \hline
\end{tabular}
}
\caption{The nuclear empirical parameters values and their variations at saturation density for Meta-model, estimated from the analyses of various experimental data. These uncertainties are taken from~\cite{JM1}.}
\label{tab:MM_params_uncertainties}
\end{table*}

\section{Results}
\label{sec:results}

From the sensitivity study, it is clear which nuclear parameters influence the different NS osbervables. However, as the parameters are varied independently, it is not evident whether there may be underlying correlations among the parameters themselves. 

\subsection{Correlation study using RMF}
\label{sec: corr_study_RMF}

In the next step, we vary all the nuclear empirical parameters within their uncertainty range from the analysis of various nuclear experiments and impose the constraints from nuclear theory and astrophysical observations, as discussed in Section~\ref{sec:Imposing_constraints}. From the posteriors, we then extract the underlying correlations among the parameters and with the NS observables. For the RMF model, we use Table \ref{tab:RMF_priors} for the values of all the nuclear parameters and their maximum possible variations, and reproduce the results from our previous works~\cite{Ghosh2022a, Ghosh2022b}. \\

\begin{table}[htbp]
\resizebox{\linewidth}{!}{
\begin{tabular}{ c c c c c c c} 
 \hline \hline
  Model & $n_{sat}$ & $E_{sat}$ & $K_{sat}$  & $J_{sym}$ & $L_{sym}$ & $m^*_D/m$ \\
& $(fm^{-3})$ & $(MeV)$ & $(MeV)$ & $(MeV)$  & $(MeV)$ &  \\ [0.5ex] 
 \hline
 RMF & 0.14-0.17 & -16$\pm$0.2 & 200-300 & 28-34 & 40-70  & 0.55-0.75 \\ [1ex]
 \hline  \hline
\end{tabular}
}
\caption{The prior range of the nuclear empirical parameters considered for the correlation study within RMF framework. This table is taken from our previous work~\cite{Ghosh2022a}.}
\label{tab:RMF_priors}
\end{table}

We vary all the parameters randomly within their uncertainty range to obtain a flat prior. For these randomly generated parameters set, we calculate the energy per particle (binding energy) as a function of baryon number density ($n_B$) for pure neutron matter (PNM). We impose the $\chi$EFT constraint (Drischler et al.~2016~\cite{drischler2016_PRC}) on the prior distribution for the baryon density range $0.08-0.21\;fm^{-3}$ and accept only the nuclear parameter sets that generate the PNM EoSs within this band for this above-mentioned baryon density range. \\

\begin{figure}[htbp]
    \centering
    \includegraphics[width = \linewidth]{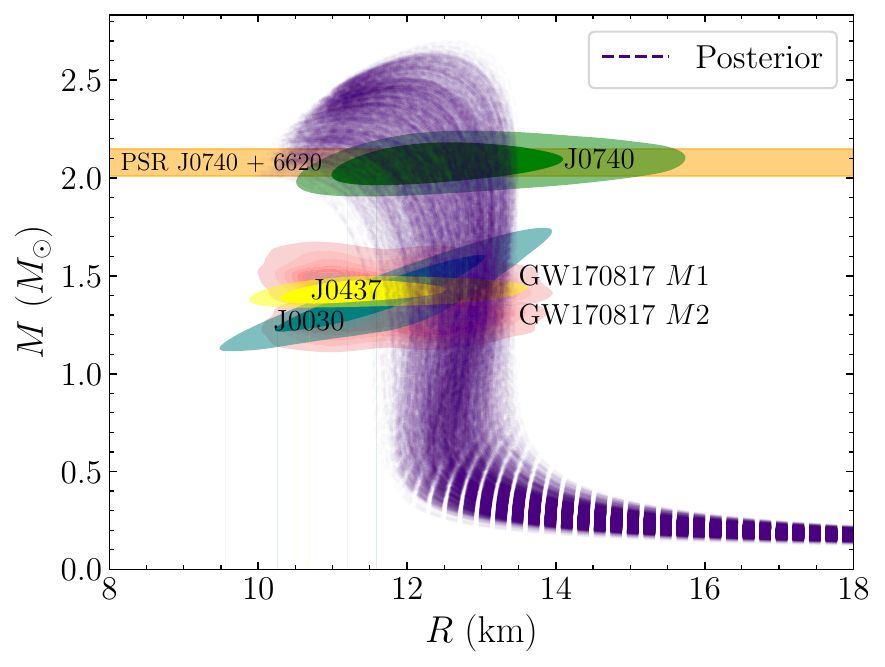}
    \caption{Mass-radius relations for posterior ANM EoSs within RMF framework after applying $\chi$EFT (Drischler et al.~2016~\cite{drischler2016_PRC})  and Astro filters. The $68\%$ and $95\%$ contours of the NICER mass-radius measurements are shown by dark and light regions, respectively~\cite{Rutherford_2024,Salmi_2024,Vinciguerra_2024,Choudhury_2024}. (See text for more details)}
    \label{fig:MR_posterior_RMF}
\end{figure}

Next, we calculate the ANM EoS for the $\chi$EFT filtered nuclear parameters sets. Following the formalism discussed in Section~\ref{sec:structure} and \ref{sec:oscillation_modes}, we obtain the mass-radius relationship, tidal deformability, and $f-$mode frequency. Now we apply the Astro constraints at high density described in section \ref{sec:Astro_constraint}. The maximum mass predicted by any EoS should exceed the mass of PSR J0740+6620. The maximum radius and dimensionless tidal deformability limits come from GW170817 for $1.4M_{\odot}$ NS. \\

\begin{figure*}[hb]
    \centering
    \includegraphics[scale = 0.52]{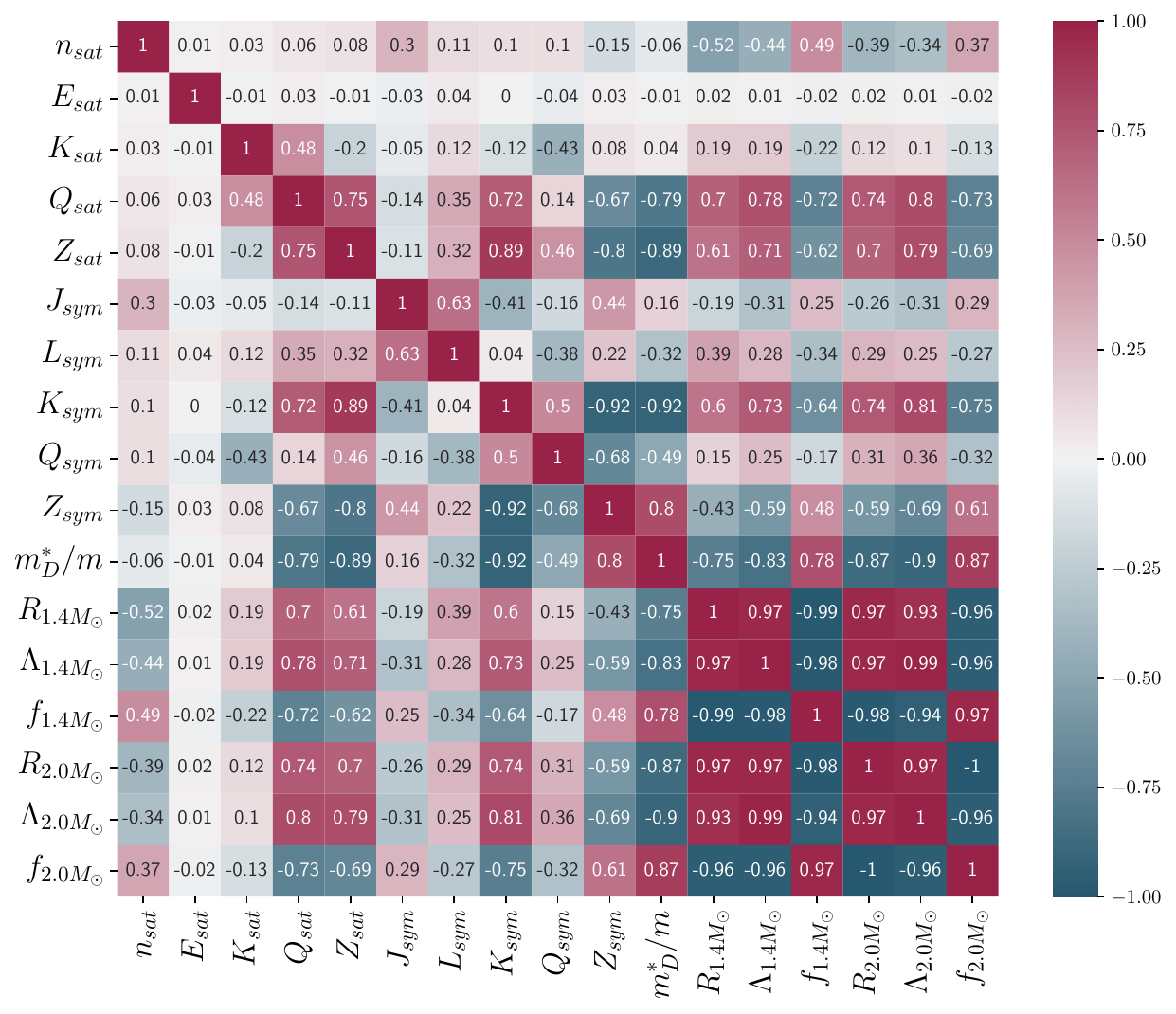}
    \caption{Pearson's linear correlations among nuclear parameters and NS observables within RMF framework after applying the $\chi$EFT (Drischler et al.~2016~\cite{drischler2016_PRC}) filter. The (negative)positive values indicate the (anti-)correlation. }
    \label{fig:RMF_xEFT_corr}
\end{figure*}

\begin{figure*}[htbp]
    \centering
    \includegraphics[scale = 0.52]{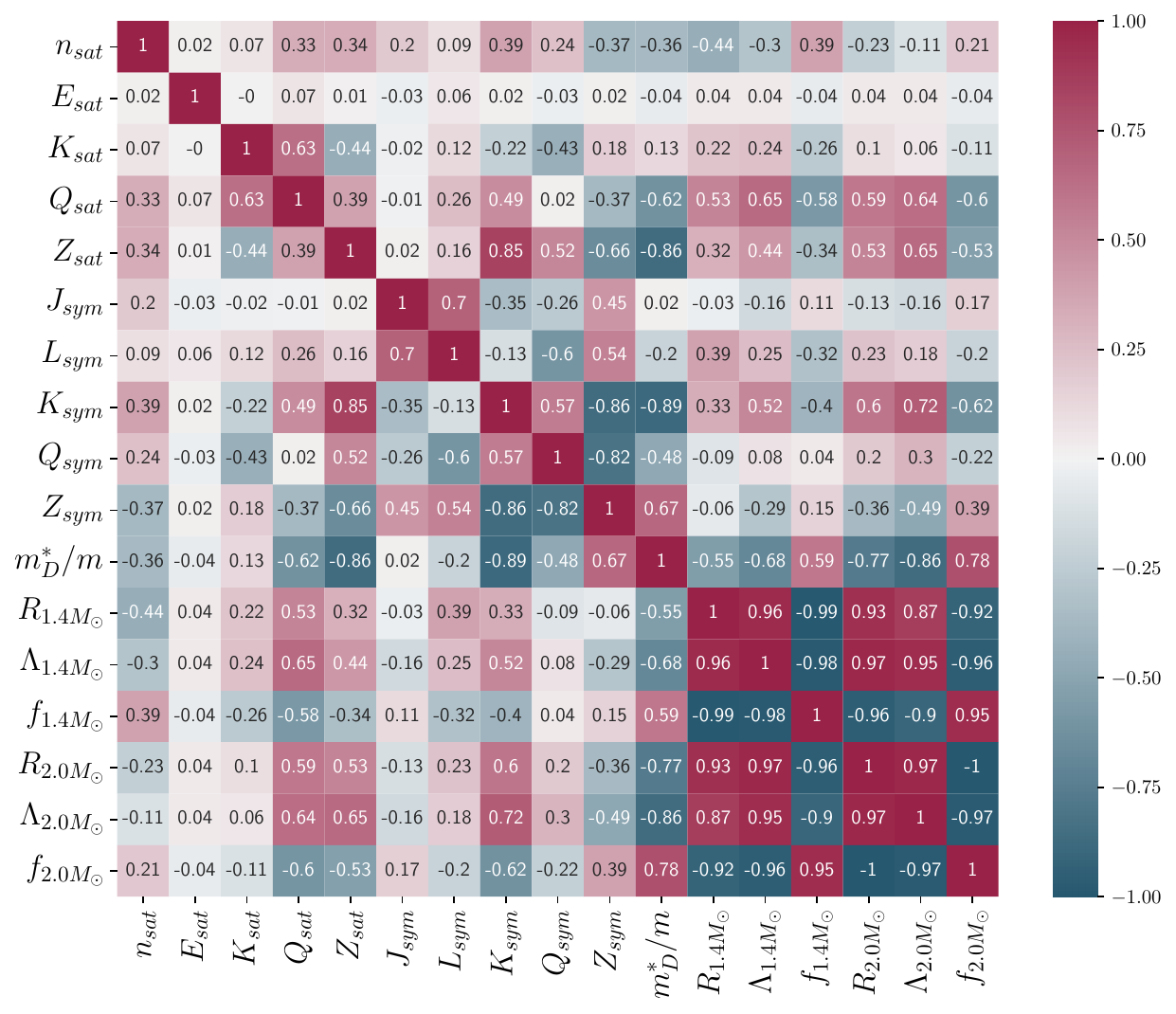}
    \caption{Same as Figure~\ref{fig:RMF_xEFT_corr}, but after applying both $\chi$EFT (Drischler et al.~2016~\cite{drischler2016_PRC}) and Astro (lower limit of maximum mass $2.01M_\odot$ from PSR J$0740 + 6620$ and upper limit on tidal deformability $\Lambda_{1.4M_\odot} < 720$ from GW170817 event) filters. }
    \label{fig: RMF_xEFT_Astro_corr}
\end{figure*}

We plot the mass-radius relations for the posterior ANM EoSs after applying these constraints in Figure \ref{fig:MR_posterior_RMF}. The mass-radius posterior is compatible with the M-R contours (shown as red patches) of the binary component from the GW170817 event, as well as the $2$-$\sigma$ contours of the NICER mass-radius measurement of the observed pulsars, as shown in Figure~\ref{fig:MR_posterior_RMF}. The fraction of EoSs that pass only the $\chi$EFT constraint is around $30\%$, while only around $21\%$ EoSs pass both filters. We generate 10000 random parameter sets within the uncertainty range as priors, out of which we effectively obtain around 2100 parameter sets that pass both filters. We obtain the correlation matrices shown in Figures~\ref{fig:RMF_xEFT_corr} and \ref{fig: RMF_xEFT_Astro_corr} following Section~\ref{sec:Pearson_corr} after applying only the $\chi$EFT filter and both $\chi$EFT + Astro filters, respectively. 
\\ 

From both the correlation matrices, we observe that the nuclear parameters $J_{sym}$ and $L_{sym}$ are strongly correlated (0.7 after passing both filters). The parameter $n_{sat}$ shows a weak correlation with $J_{sym}$ in both matrices. The anti-correlation between $n_{sat}$ and $m^*_D/m$ is moderate after applying both filters (-0.36), whereas it shows no correlation when it passes only the $\chi$EFT filter.
The parameters $n_{sat}$ and $L_{sym}$ have moderate correlations with $1.4M_{\odot}$ NS observables ($R_{1.4M_\odot}$, $\Lambda_{1.4M_\odot}$, $f_{1.4M_\odot}$). But they have a very weak correlation for $2M_{\odot}$ NS observables ($R_{2.0M_\odot}$, $\Lambda_{2.0M_\odot}$, $f_{2.0M_\odot}$). 
All the NS observables are strongly correlated with each other, which is expected from the equation (\ref{eq:lamda_eq}). The nuclear parameter $m^*_D/m$ is strongly correlated with NS observables ($R$, $\Lambda$, and $f-$mode frequency). This correlation further increases with increasing NS masses. Therefore, in the RMF model, the Dirac effective nuclear mass ($m^*_D/m$) plays a crucial role in determining the NS observables like mass, radius, and dimensionless tidal deformability. These results are consistent with our previous works~\cite{Ghosh2022a,Pradhan2021,Pradhan2022}. While NS observables are controlled by the EoS above saturation primarily via the effective mass in the RMF model, in other nuclear models (e.g. in the meta-model) they may be controlled via higher-order parameters. To improve comparability, we calculate the higher-order parameters ($Q_{sat}, Z_{sat}, K_{sym}, Q_{sym}, Z_{sym}$) by taking the higher-order derivatives of the energy per baryon with respect to baryon density for each set of posteriors (following~\cite{Chen_2014}) and include them in the correlation matrices in Figures~\ref{fig:RMF_xEFT_corr} and \ref{fig: RMF_xEFT_Astro_corr}.
Note that, these higher-order parameters are not varied independently; rather, they are calculated from the lower-order nuclear parameters ($n_{sat},E_{sat},K_{sat},J_{sym},L_{sym},m^*_D/m$) varied within their allowed uncertainties as given in Table~\ref{tab:RMF_priors}.
For the calculated higher-order parameters ($Q_{sat},Z_{sat},K_{sym},Q_{sym},Z_{sym}$), correlations are observed with the lower-order nuclear parameters ($K_{sat},J_{sym},L_{sym}$) as well as among themselves. One finds that $Q_{sat}, Z_{sat}$, and $K_{sym}$ exhibit strong correlations with the Dirac effective mass as well as with NS observables on application of the $\chi$EFT filter, which are weakened on application of the Astro filter. There is a moderate correlation between $Z_{sym}$ and NS observables, while the $Q_{sym}$ always displays a weak correlation. \\

 \begin{figure*}[htbp]
    \centering
    \includegraphics[scale = 0.68]{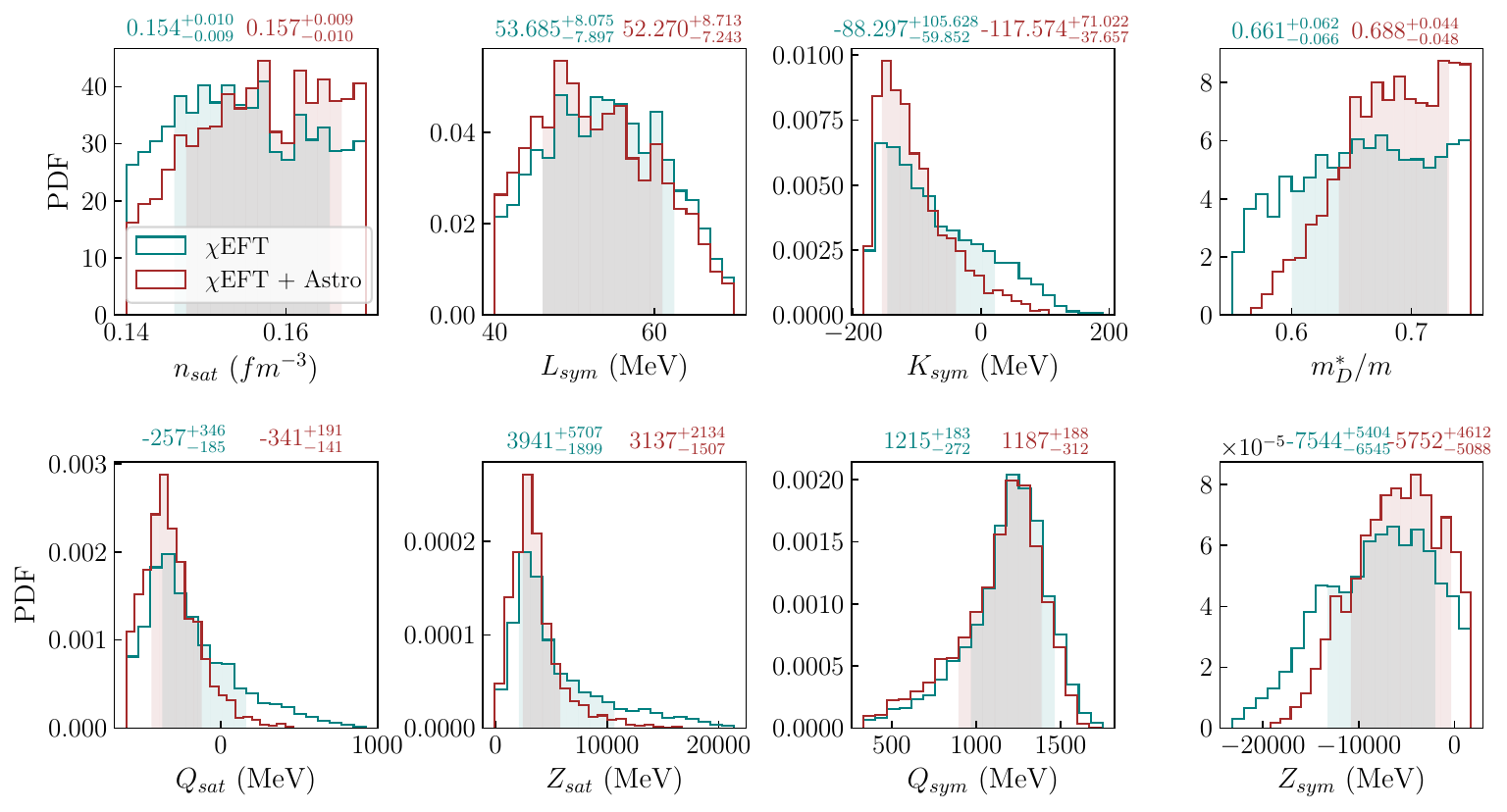}
    \caption{The probability density functions (PDFs) of the most sensitive parameters in the RMF model ($n_{sat}$, $L_{sym}$ and $m^*_D/m$) along with the calculated higher-order parameters, after passing both $\chi$EFT (Drischler et al.~2016~\cite{drischler2016_PRC}) and Astro filters. The mean and 1$\sigma$ values (shown as filled regions) of each parameter are provided on the top of their PDF.}
    \label{fig: PDF_RMF_params}
\end{figure*}

The probability density functions (PDFs) of some of the nuclear empirical parameters ($n_{sat}$, $L_{sym}$, $m^*_D/m$) which are strongly correlated with NS observables along with the calculated higher-order parameters ($Q_{sat},Z_{sat},K_{sym},Q_{sym},Z_{sym}$) are shown in Figure~\ref{fig: PDF_RMF_params}. From the PDFs, one can observe that the symmetry energy slope parameter ($L_{sym}$) gets slightly constrained. Another important observation is that the inclusion of the Astro filter favours the higher values of the Dirac effective mass. The calculated values of the higher-order nuclear parameters for the posteriors are found to span a wide range.

\subsection{Correlation study using MM}

Here, we follow similar steps as in RMF for the correlation study. We keep the same prior ranges for the lower-order parameters ($n_{sat}$, $E_{sat}$, $K_{sat}$, $J_{sym}$, $L_{sym}$), as these are also present in RMF model. The prior ranges from the higher-order parameters ($K_{sym}$, $Q_{sat}$, $Z_{sat}$, $Q_{sym}$, $Z_{sym}$), Landau effective mass ($m^*_L/m$), and isospin mass splitting ($\Delta m^*_L/m$) are taken from~\cite{JM1}. The large uncertainties used as priors for the correlation study within Meta-model framework are given in Table~\ref{tab:MM_priors}.

\begin{table*}[!h]
\centering
\resizebox{1 \textwidth}{!}{
\begin{tabular}{ c c c c c c c c c c c c c} 
 \hline \hline
Model & $n_{sat}$ & $E_{sat}$ & $K_{sat}$ & $Q_{sat}$ & $Z_{sat}$ & $J_{sym}$ & $L_{sym}$ & $K_{sym}$ & $Q_{sym}$ & $Z_{sym}$ & $m^*_L/m$ & $\Delta m^*_L/ m$\\
 & $(fm^{-3})$ & (MeV) & (MeV) & (MeV) & (MeV) & (MeV) & (MeV) & (MeV) & (MeV) & (MeV) & & \\  [0.5ex]
 \hline 
 MM & 0.14-0.17 & -16 $\pm$ 0.2 & 200-300 & 300 $\pm$ 400 & -500 $\pm$ 1000 & 28-34  & 40-70 & -100 $\pm$ 100 & 0 $\pm$ 400 & -500 $\pm$ 1000 & 0.65-0.85 & 0.0-0.2\\ [1ex] \hline \hline
\end{tabular}
}
\caption{The prior range of the nuclear empirical parameters considered for the correlation study within the Meta-model framework.}
\label{tab:MM_priors}
\end{table*}

For the correlation study in the Meta-model, we use $\chi$EFT constraint (Drischler et al.~2016~\cite{drischler2016_PRC}) at low density and the lower limit of maximum mass constraint of $2.01 M_{\odot}$(PSR J0740+6620), and the maximum tidal deformability of $1.4M_{\odot}$ NS ($\Lambda_{1.4M_\odot} < 720$ from GW170817) at higher density. We first apply the $\chi$EFT filter on the randomly generated uniform prior PNM EoSs between density region $0.08-0.21\;fm^{-3}$ within the Meta-model framework. Following the discussion in Section~\ref{sec:structure} and \ref{sec:oscillation_modes}, we calculate the NS observables. Then applying the Astro filter, we obtain the posterior in this Meta-model. Here, we generated $12500$ priors in this model. Among those only $3000$ ($24\%$) EoSs pass the $\chi$EFT filter, and around $1950$ ($15.6\%$) EoSs pass both the filters ($\chi$EFT $+$ Astro).\\

  \begin{figure}[!h]
    \centering
    \includegraphics[width=\linewidth]{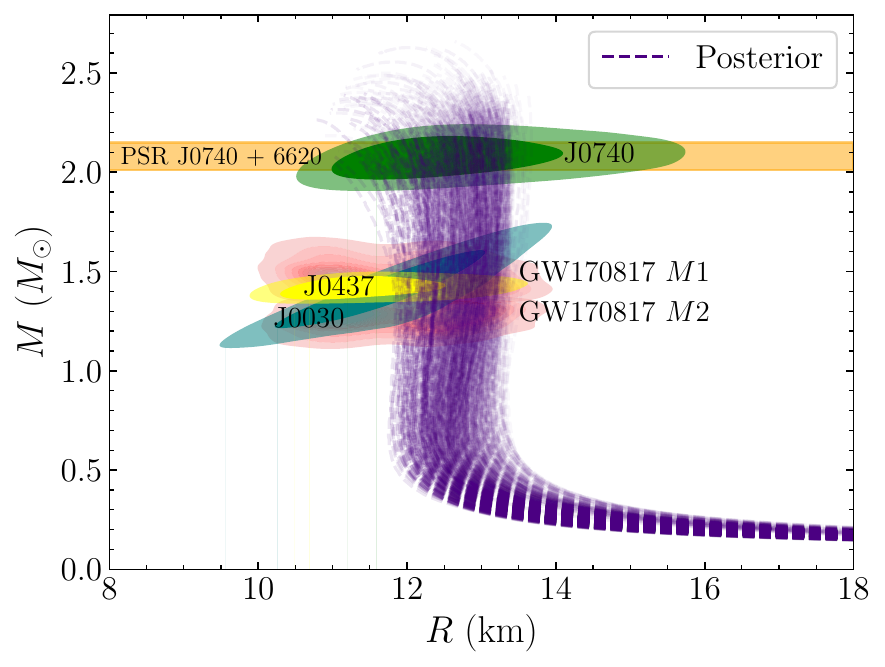}
    \caption{Mass-radius posteriors for ANM EoSs within Meta-Model framework after applying both $\chi$EFT (Drischler et al.~2016~\cite{drischler2016_PRC}) and Astro filters. The $68\%$ and $95\%$ contours of the NICER mass-radius measurements are shown by dark and light regions, respectively~\cite{Salmi_2024,Vinciguerra_2024,Choudhury_2024,Rutherford_2024}.}
    \label{fig:MR_posterior_MM}
\end{figure}

The M-R relations for the EoS posteriors have been plotted in Figure~\ref{fig:MR_posterior_MM} after applying both $\chi$EFT and Astro constraints. We observe that the posterior mass-radius curves are compatible with the $2$-$\sigma$ contours of the recent NICER observational data for pulsars J0740, J0030, and J0437~\cite{Rutherford_2024}. The posterior mass-radius relations are also compatible with $M$-$R$ contours of GW170817 event. \\

\begin{figure*}[htbp]
    \centering
    \includegraphics[scale = 0.52]{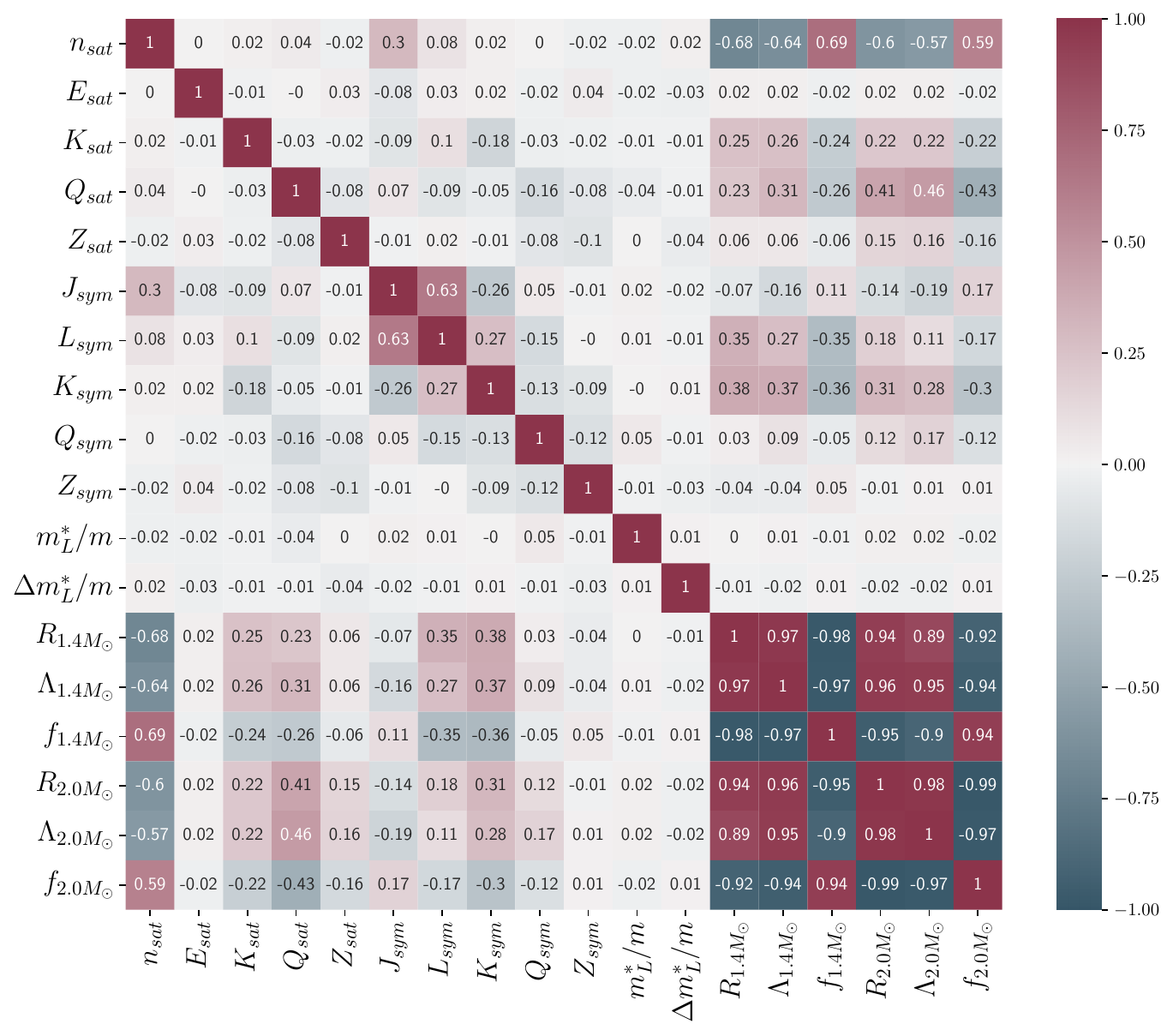}
    \caption{Correlations among the nuclear parameters and NS observables within Meta-Model framework after applying only $\chi$EFT (Drischler et al.~2016~\cite{drischler2016_PRC}) filter. The (negative)positive values indicate the (anti-)correlation.}
    \label{fig: MM_xEFT_corr}
\end{figure*}

\begin{figure*}[htbp]
    \centering
    \includegraphics[scale = 0.52]{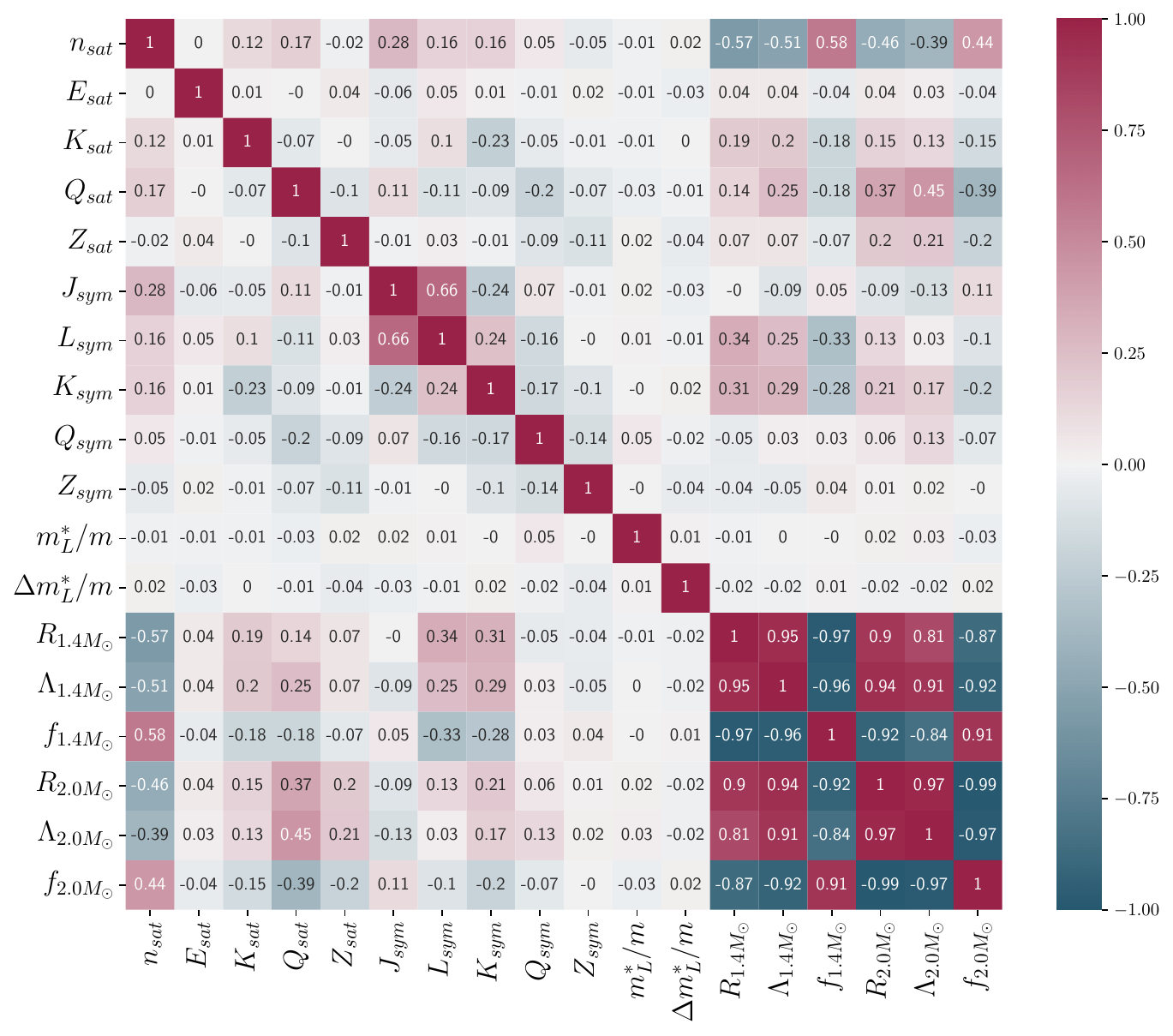}
    \caption{Same as Figure~\ref{fig: MM_xEFT_corr}, but after applying both $\chi$EFT (Drischler et al.~2016~\cite{drischler2016_PRC}) and Astro (lower limit of maximum mass $2.01M_\odot$ from PSR J$0740 + 6620$ and upper limit on tidal deformability $\Lambda_{1.4M_\odot} < 720$ from GW170817 event) filters.}
    \label{fig: MM_xEFT_Astro_corr}
\end{figure*}

\begin{figure*}[hb]
    \centering
    \includegraphics[scale = 0.65]{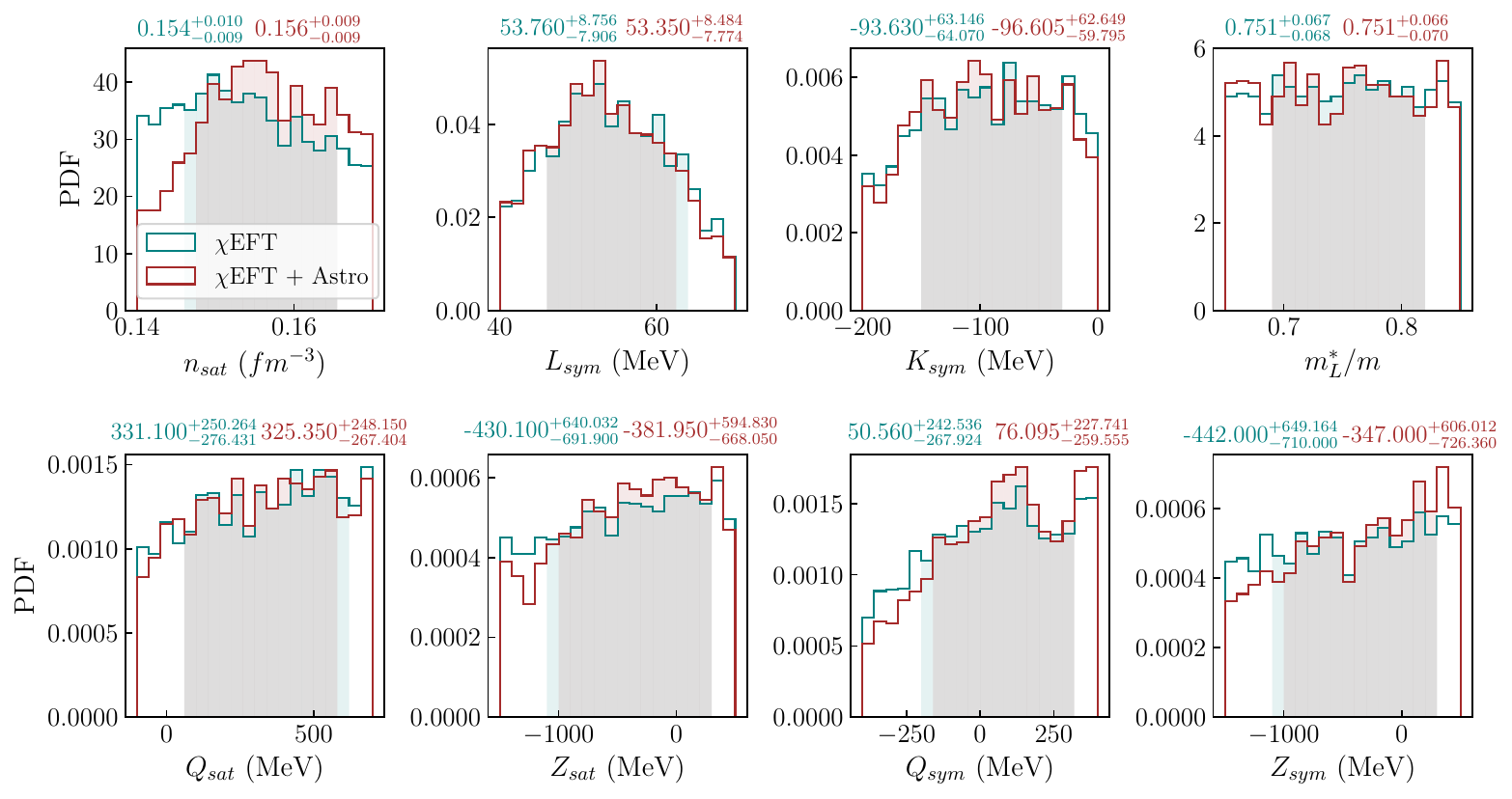}
     \caption{The probability density functions (PDFs) of the sensitive nuclear parameters in the Meta model after applying both $\chi$EFT (Drischler et al. 2016~\cite{drischler2016_PRC}) and Astro filters. The mean and 1$\sigma$ values (shown as filled regions) of each parameter are provided on the top of their PDF.}
    \label{fig: PDF_MM_HOparams}
\end{figure*}

\begin{figure*}[hb]
    \centering
    \includegraphics[scale = 0.69]{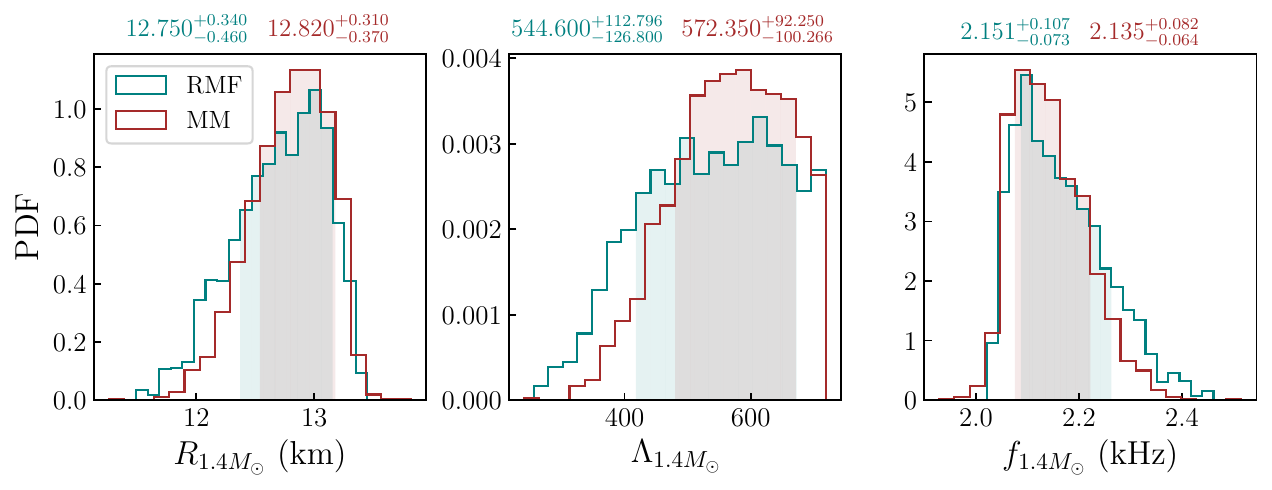}
    \caption{Comparison between the probability density functions (PDFs) of the NS observables ($R_{1.4 M_\odot}$, $\Lambda_{1.4 M_\odot}$, $f_{1.4 M_\odot}$) in the RMF and Meta-model after passing both $\chi$EFT (Drischler et al. 2016~\cite{drischler2016_PRC}) and Astro filters. The mean and 1$\sigma$ values (shown as filled regions) of each observable are provided on the top of their PDF.}
    \label{fig: PDF_observables_comparison}
\end{figure*}

We obtain the correlation matrix among nuclear parameters and NS observables for MM after following the description in Section~\ref{sec:Pearson_corr}. The correlation matrices after applying only $\chi$EFT filter and after applying both the filters are shown in Figures~\ref{fig: MM_xEFT_corr} and \ref{fig: MM_xEFT_Astro_corr} respectively. The correlation matrix for MM shows that the parameters $J_{sym}$ and $L_{sym}$ are strongly correlated ($0.66$ after passing both filters). The parameters $L_{sym}$ and $K_{sym}$ are weakly correlated in both matrices. They show a moderate correlation with the NS observables of $1.4 M_{\odot}$. This correlation decreases for $2M_{\odot}$ NS observables. $Q_{sat}$ is moderately correlated with the NS observables of mass $2M_\odot$. It is expected as $Q_{sat}$ affects the high-density behaviour much compared to the other higher-order parameters. There is a weak correlation between $n_{sat}$ and $J_{sym}$ $(0.3)$ after applying only $\chi$EFT filter, and it decreases after applying the Astro filter. Strong correlations are observed between $n_{sat}$ and NS observables ($R$, $\Lambda$, and $f-$mode frequency) after applying $\chi$EFT filter. However, these correlations become moderate after applying the Astro filter. The $f-$mode frequency shows comparatively strong correlation with $n_{sat}$ than the radius and tidal deformability. The Landau effective mass ($m^*_L/m$) and its isospin mass splitting ($\Delta m^*_L/m$) at saturation are not correlated with the other nuclear parameters as well as with the NS observables. Note that, we only admit those EoSs in the posterior that simultaneously satisfy the minimal physics conditions (positive symmetry energy and causality for all densities up to $n_{max}$, the central density corresponding to the heaviest observed neutron star with mass $M=2.01M_\odot$) along with the aforementioned constraints. For the posterior, we have checked that the value of $n_{max}$ typically lies in the range $3-4n_{sat}$. 
\\

The higher-order parameters in the Meta-model exhibit large uncertainties. To assess how much the nuclear parameters get constrained, we present the probability density functions (PDFs) of the posterior nuclear parameters ($n_{sat}$, $L_{sym}$, $K_{sym}$, $Q_{sat}$, $Z_{sat}$, $Q_{sym}$, and $Z_{sym}$) in Figure~\ref{fig: PDF_MM_HOparams}, which govern the high-density behaviour in this model and thus influence the NS observables. We also include the PDF of the Landau effective mass ($m^*_L/m$) for comparison with the Dirac effective mass ($m^*_D/m$). From the PDFs distributions, we observe that the symmetry energy slope parameter $L_{sym}$ is constrained better than the other parameters in this model. The higher-order parameters ($K_{sym}$, $Q_{sat}$, $Z_{sat}$, $Q_{sym}$, and $Z_{sym}$) are very poorly constrained. The PDFs of the $\chi$EFT and $\chi$EFT $+$ Astro posteriors do not differ significantly. However, we obtain a flat posterior for the Landau effective mass as it is not correlated with the NS observables. We compare the posterior radius, tidal deformability, and $f-$ mode frequency of $1.4M_\odot$ Neutron Star between the RMF and Meta-models in Figure~\ref{fig: PDF_observables_comparison}. The Meta-model predicts a slightly larger radius and tidal deformability, as well as a slightly lower $f$-mode frequency for 1.4$M_\odot$ neutron stars, compared to the RMF model.




\subsection{Improved constraints}

\begin{figure*}[htbp]
    \centering
    \includegraphics[scale = 0.52]{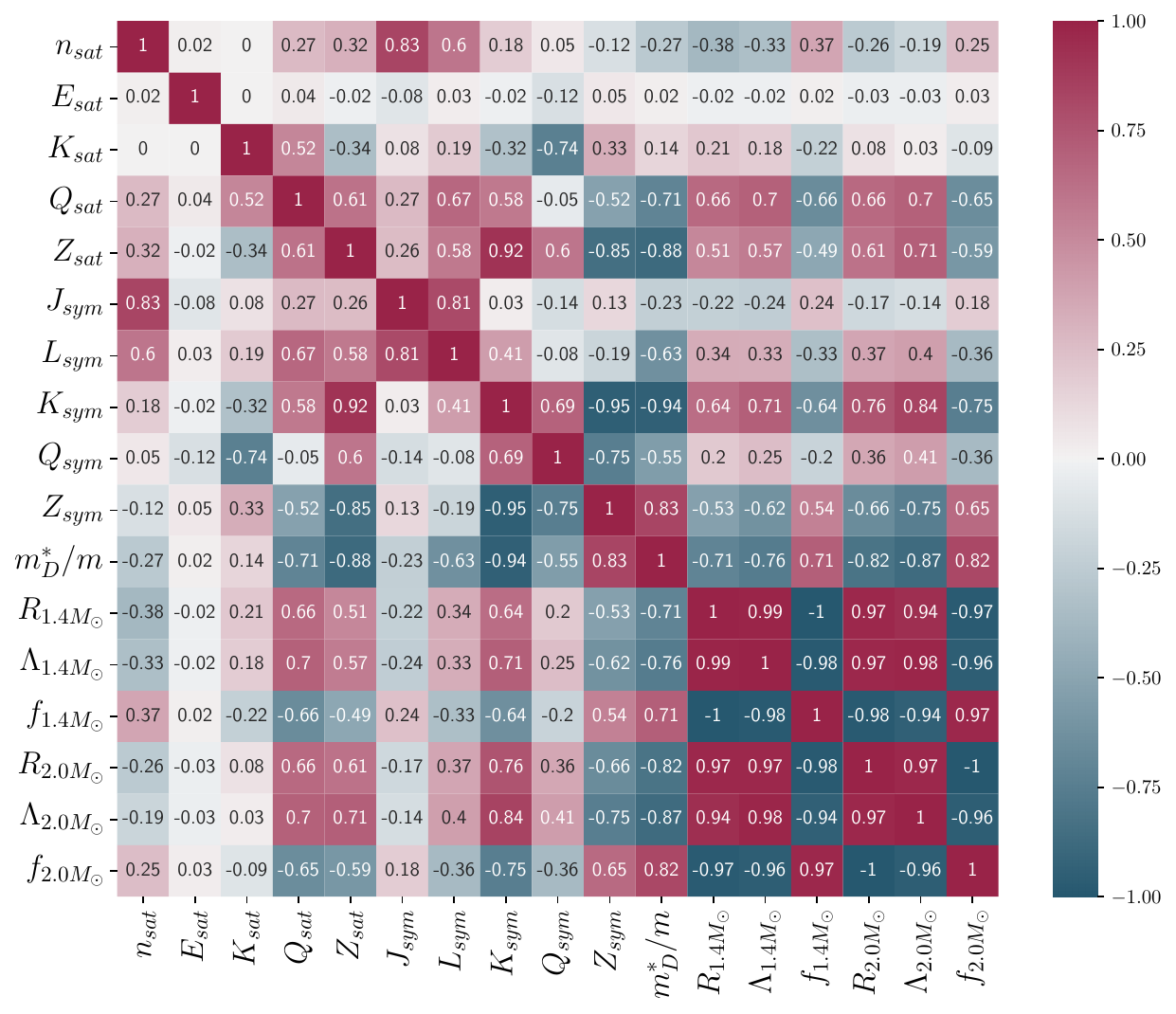}
    \caption{Correlations among nuclear parameters and NS observables within RMF framework after applying improved $\chi$EFT (Keller et al. 2023~\cite{Keller_Hebeler_2023prl}) filter for pure neutron matter (PNM).}
    \label{fig:corr_newxEFT_rmf}
\end{figure*}

We presented our results of the correlation study in Section~\ref{sec: corr_study_RMF} using the $\chi$EFT constraint for PNM by Drischler et al.~2016~\cite{drischler2016_PRC} within RMF framework. In this section, we investigate whether the recently extended $\chi$EFT (up to $0.25\;fm^{-3}$) calculation (Keller et al.~2023~\cite{Keller_Hebeler_2023prl}) for PNM can improve the previous results (Figure~\ref{fig:xEFT_pnm_band_comparison} compares two $\chi$EFT bands of PNM). In order to do that, we follow the similar steps as discussed in Section~\ref{sec: corr_study_RMF}. The correlation matrix between nuclear parameters and NS observables within RMF framework after applying only improved $\chi$EFT (Keller et al.~2023~\cite{Keller_Hebeler_2023prl}) filter for PNM is shown in Figure~\ref{fig:corr_newxEFT_rmf}.\\

\begin{figure}[htbp]
    \centering
    \includegraphics[width = \linewidth]{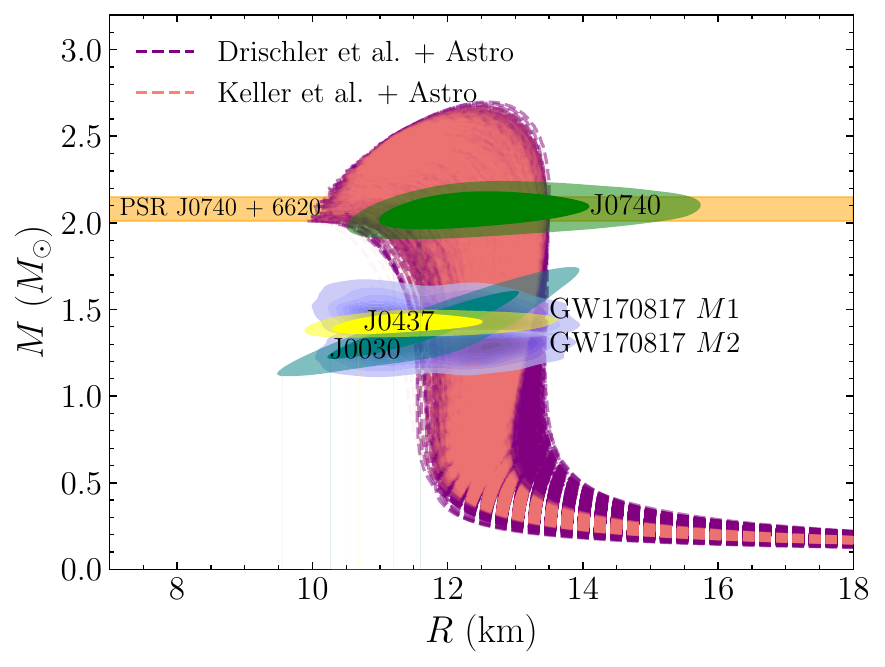}
    \caption{The comparison between the mass-radius posteriors of the previous $\chi$EFT (Drischler et al. 2016~\cite{drischler2016_PRC}) + Astro and the improved $\chi$EFT (Keller et al. 2023~\cite{Keller_Hebeler_2023prl}) + Astro constraints within RMF framework. The $68\%$ and $95\%$ contours of the NICER mass-radius measurements are shown by dark and light regions (`teal', `green', and `yellow') respectively~\cite{Salmi_2024,Vinciguerra_2024,Choudhury_2024,Rutherford_2024}. The blue M-R contours are for the binary component of the GW$170817$ event.}
    \label{fig:mr_comparison}
\end{figure}

If we compare Figures~\ref{fig:RMF_xEFT_corr} and \ref{fig:corr_newxEFT_rmf}, we can see that the strong nuclear correlation between $J_{sym}$ and $L_{sym}$ increases (from $0.63$ to $0.81$). The moderate correlations between $n_{sat}$ and NS observables become weak after using new $\chi$EFT constraint (Keller et al.~2023~\cite{Keller_Hebeler_2023prl}). The correlations of $n_{sat}$ with $J_{sym}$ and $L_{sym}$, which were weak in the previous result, get stronger after using this improved $\chi$EFT constraint. The weak anti-correlation between $L_{sym}$ and $m^*_D/m$ gets increased (from -0.32 to -0.63). As the $\chi$EFT constrains the low-density region, the correlations between $m^*_D/m$ and NS observables do not change significantly. The correlations of the calculated higher-order parameters in the RMF model, among themselves as well as with the NS observables, have been included for completeness. We show a comparison between the mass-radius posteriors obtained after imposing the previous $\chi$EFT~\cite{drischler2016_PRC} + Astro and the new $\chi$EFT~\cite{Keller_Hebeler_2023prl} + Astro constraints in Figure~\ref{fig:mr_comparison}. We observe that the improved $\chi$EFT of PNM provides tighter constraints on the radius of low-mass Neutron Stars. \\

\section{Discussions}
\label{sec:discussions}

\subsection{Summary of this work}
\label{sec:summary}

In a series of works based on a formalism inspired by the Bayesian scheme~\cite{Ghosh2022a,Ghosh2022b,Pradhan_zeta_2023}, information from nuclear theory and state-of-the-art multi-messenger astrophysical data were imposed to constrain the parameter space of uncertainties in nuclear empirical parameters within the RMF model framework with non-linear couplings. A study of the correlations revealed the Dirac effective nucleon mass to be the most dominant parameter that controls NS observable properties (radius, tidal deformability and $f$-mode frequency). It was also demonstrated how this information can be useful for constraining the nuclear parameters, in particular the nucleon effective mass~\cite{Pradhan_2023,Pradhan_2024}, via NS asteroseismology with future $f$-mode observations. In this work, our main motivation is to determine whether such a correlation is generic or an artifact of the underlying nuclear (non-linear RMF) model considered. 
\\

To investigate this, we first performed a sensitivity study to determine the important nuclear parameters that influence the NS observables. We reproduced the results of the correlation study within the RMF scheme, following~\cite{Ghosh2022a}.  
To test the model dependence of the correlations, we then repeated the sensitivity and correlation studies using a different nuclear EoS model, the generic Meta-model. By construction, the meta-model is agnostic and does not assume any underlying nuclear model; therefore, it can be used to check whether correlations are physical or spurious. The caveats of the MM are that it is non-relativistic (therefore becomes acausal at high densities) and is based on an expansion valid close to the saturation density. At large densities, the higher order parameters in the MM, which are highly uncertain, gain importance and therefore the validity of the model at densities away from saturation becomes questionable.\\

As a summary of our comparative study considering the two models, we observed that the strong correlation between the symmetry energy $J_{sym}$ and its slope $L_{sym}$ is present in both models and therefore this correlation is intrinsic. The moderate correlation between $L_{sym}$ and NS observables is also present in both models. The saturation density $n_{sat}$ shows a moderate correlation with NS observables in both models (comparatively stronger in MM than in the RMF). The moderate correlation of $K_{sym}$ and $Q_{sat}$ with the NS observables are found in the MM, but these higher-order parameters are not present in the RMF model. A large disagreement in the correlation has been seen for the effective nucleon mass with NS observables and with $n_{sat}$. In the RMF model, we recover the strong correlation of the Dirac effective mass $m^*_D/m$ with radius and dimensionless tidal deformability of NS, which increases with increasing NS masses. In contrast, the Landau effective mass $m^*_L/m$ in MM is not at all correlated with NS observables.\\

From the results of this study, we come to the conclusion that most of the correlations between nuclear parameters and NS observables are common to both models (and physical), except for the effective nucleon mass with NS observables. However one must note that the definition of the effective nucleon mass varies across models, and is different in the case of the RMF and Meta-model. In non-linear RMF models, the effective nucleon mass is known to control the high density behaviour of the EoS and is therefore correlated with the observables.  It was demonstrated in Pradhan et~al.~(2023)~\cite{Pradhan_zeta_2023} that, upon adding the vector self-interaction coupling ($\zeta$), which also influences the high density behaviour of the EoS, the correlation of the NS observables with the nucleon effective mass is reduced. In the MM, the high density EoS behaviour in this model is controlled by higher order parameters such as $K_{sym},Q_{sat},Z_{sat},Q_{sym}$ and $Z_{sym}$. The Landau effective mass $m_L^*/m$ therefore shows no correlation with the NS observables. However these higher order parameters have large uncertainties as they cannot be constrained by nuclear experimental data close to saturation density. Based solely on the results from the meta-model, we cannot argue that the effective mass would not have any correlation with observables in general. Rather, we found the correlation to be model-dependent: it appears in the RMF model but is absent in the agnostic meta-model. This behavior depends on how the effective mass is defined within a given model.\\

In the present work, we mostly focused on the model dependence of the nuclear correlations. It was demonstrated by several recent works within the RMF formalism that the effective nucleon mass affects the $f$-mode characteristics in NSs and therefore future $f$-mode detections can be used to constrain its value \cite{Pradhan2021,Pradhan_2023,Pradhan_2024}. As an extension of this work, one may investigate whether higher order nuclear parameters could similarly be constrained from future NS observations. We found that if we impose constraints from $\chi$EFT calculations, current astrophysical data from NS maximum mass, tidal deformability of GW170817, and NICER mass-radius measurements to 2-$\sigma$ precision, then the values of the higher order parameters are not significantly constrained. We checked that current mass-radius measurements to a 1-$\sigma$ precision improve constraints on the high-density behavior of
nuclear symmetry energy ($L_{sym}$, $K_{sym}$) but are unable to significantly constrain higher-order parameters. It was suggested in~\cite{Li_2024} employing the Meta-model that the more precise measurements of the NS radius (1.0 - 0.1 km) may significantly help to constrain the parameter space as well as the pairwise correlations among the nuclear parameters.
\\


\subsection{Comparison with other works}

Recent works (see e.g.~\cite{Dinh_Thi_2021,Mondal_Gulminelli_2023}) employed the nuclear meta-modelling technique to investigate correlations among the nuclear empirical parameters, by combining NS global properties as well as measured properties of nuclei (such as binding energy, charge radii and neutron skin thickness) within a full Bayesian analysis. They found that at low densities, information from nuclear experiments imposes strong constraints on the lower order (up to second order) isoscalar parameters, while adding subsequent information from $\chi$EFT, neutron skin data and astrophysical observations marginally affect the posterior distributions of the parameters. The isovector parameters were found to be poorly constrained by nuclear physics experiments but were strongly affected by the $\chi$EFT constraints. The low density filters were also found to have a non-negligible impact on the higher order parameters $Q_{sat}$ and $K_{sym}$. The parameters $Z_{sat}$ and $Z_{sym}$ did not have any correlations with other parameters and were unaffected by the different constraints. The constraints from astrophysical observables were seen to play an important role on the parameters at high density, as expected. However, more precise astrophysical data or from heavy-ion collision experiments are required to tighten the constraints on higher-order parameters. In this study, we considered a broader prior range for the saturation density ($n_{sat}$) compared to the work by Dinh Thi et al.~\cite{Dinh_Thi_2021}. As a result, we observed relatively stronger correlations between $n_{sat}$ and neutron star properties compared to their findings~\cite{Dinh_Thi_2021}, reflecting the significant influence of $n_{sat}$ on these properties. However, it is important to note that $n_{sat}$ is well constrained by data from low-density experiments~\cite{Klausner_2024} and this correlation does not help to improve the information. Once $n_{sat}$ is fixed, more precise astrophysical measurements can be used to constrain other key parameters, such as $Q_{sat}$, $L_{sym}$, and $K_{sym}$ which are then more strongly correlated. It is also confirmed that radius measurements from NICER~\cite{Miller_2019,Riley_2019,Miller_2021,Riley_2021} do not show a significant impact on the EoS, in agreement with other works~\cite{pang2021,Raaijmakers2021}. 
These works also noted a moderate (anti-)correlation with ($E_{sat}$) $n_{sat}$. The isovector parameters $J_{sym}$, $L_{sym}$ and $L_{sym}$, $K_{sym}$  were found to correlate strongly among themselves, while higher order parameters $K_{sat} - Q_{sat}$ and $K_{sym} - Q_{sym}$ also showed weak correlations due to narrow $\chi$EFT bands at low densities. \cite{Tsang_2024} performed Bayesian analyses of 12 nuclear experimental constraints combined with 3 complementary constraints from astrophysical observations to obtain the EoS of neutron-rich matter at number densities from $\sim 0.5 - 3 n_{sat}$ using a Meta-modelling scheme. Their results are compatible at low density with the recently developed $\chi$EFT band up to 2$n_{sat}$, but the disagreement increases with increasing density, demonstrating a tension between $\chi$EFT and nuclear experimental data.
\\

Frequencies of $f-$mode oscillations calculated within the Cowling approximation were also recently explored within the Meta-model scheme~\cite{Montefusco2024}. However this work did not perform a study of correlations of nuclear parameters with $f$-mode frequencies. In another study~\cite{Grams_2024}, the sensitivity of $f$-mode frequencies to high-order nuclear parameters was studied using the meta-model version of the chiral Hamiltonian H2. It was concluded that $K_{sym}$ impacts the frequencies for low mass stars while $Q_{sym}$ and $Q_{sat}$ affect intermediate and heavy NSs. However, one may note that in this study each nuclear parameter was varied independently and therefore underlying correlations among the parameters themselves may exist. 
\\

Other studies in the recent literature have also confronted non-relativistic (Skyrme) and non-linear relativistic mean-field models with terrestrial experimental (finite nuclei and heavy-ion collisions) and astrophysical data within the Bayesian framework~\cite{Imam_2024,Venneti_2024}. \cite{Imam_2024} observed that the EoSs constrained by the maximum mass limit were further narrowed on the inclusion of experimental data, but the impact of further addition of other astrophysical constraints is marginal. This indicates that more precise observations are necessary to further refine the EoS. They also noted a significant difference in the values of nuclear matter parameters obtained from non-relativistic vs relativistic models: the Skyrme model predicted a softer SNM EoS as well as symmetry energy at saturation. \cite{Venneti_2024} compared the distribution of nuclear matter and NS properties from the EoSs obtained by implicit and explicit treatment of finite nuclei constraints and found them to be distinctly different. Using an extended RMF model with non-linear couplings, the influence of the slope parameter $L_{sym}$ on NS observables and $f$-modes was also investigated in~\cite{Lopes_2023}. They concluded that the $L_{sym}$ parameter could be constrained from the more precise measurements of neutron skin thickness experiments and the next-generation gravitational wave detectors in the near future. A comparative study based on relativistic and non-relativistic energy-density functionals was also performed in~\cite{Zhou_2024}, to constrain the nuclear symmetry energy from low to high densities using a Bayesian approach.
\\

 In a few other works, the correlations between various EoS parameters, properties of nuclear matter and NSs have also been explored within the framework of density-dependent covariant density functional (DD-CDF) models~\cite{Malik_2022,Scurto_2024}. The posterior distributions were found to be compatible with the ones obtained by meta-modelling techniques. Imposing constraints from nuclear data, $\chi$EFT and bound from maximum NS mass, correlations among nuclear matter and NS properties were examined~\cite{Malik_2022,Beznogov_Raduta_2023}. Among the strongest correlations were found to be between $K_{sym}$, $K_{sat}$, $Q_{sat}$ and $R_{1.4}$. Only a weak correlation was found between $L_{sym}$ and $R_{1.4}$, which can be explained by the narrow domain of $L_{sym}$ considered resulting from PNM constraints. The parametrized density-dependent RMF model~\cite{Beznogov_Raduta_2023,Thapa2023} was also used to calculate $f$-mode frequencies within the Cowling approximation~\cite{Thapa2023} and a correlation study was performed. The $f$-mode frequency showed anti-correlation with $K_{sat}$, $Q_{sat}$ and $K_{sym}$, which was explained considering that the frequency scales as the average density of the star. $f_{1.4M_\odot}$ appeared to be correlated with the effective nucleon mass, in agreement with our previous works. 
\\

\subsection{Implications and outlook}
Multi-disciplinary data at different densities are providing complementary information to constrain the EoS of dense matter. The calculations from $\chi$EFT for PNM have played an important role in constraining the isovector nuclear parameters at low densities, that is otherwise not possible only from nuclear experimental data near saturation, which is relevant for SNM and therefore only constrains isoscalar properties. These calculations are now being refined and extended to higher densities and also to $\beta$-equilibrated matter relevant for NSs, thus improving our understanding of the NS EoS. Heavy-ion collision experiments also provide important information at intermediate densities, but one must exercise caution as the results are dependent on the assumed transport models~\cite{Ghosh2022a}. Multi-messenger astrophysical data can provide important input for the high density part of the EoS that is inaccessible to terrestrial experiments. The maximum mass of NSs imposes important limits on the high density behaviour of the EoS. While current data from tidal deformability is presently only restricted to one event GW170817, the number of binary NS mergers detectable will improve significantly with the upcoming observing runs of the LIGO-Virgo-KAGRA network of GW detectors and the third generation detectors such as Einstein Telescope and Cosmic Explorer.\\

In future, it is also expected that we may be able to measure characteristics (frequency and damping time) of NS oscillation modes, in particular $f$-modes, which carry important information about the high density behaviour of the NS EoS. It was also studied how a future detection of $f$-modes could be used to improve our understanding of the nuclear parameters, particularly the effective nucleon mass, within the RMF formalism~\cite{Pradhan_2023,Pradhan_2024}. However this comparative study showed that the correlation of $f$-mode characteristics with the nucleon effective mass is model-dependent; it has been found in non-linear and density dependent RMF models, but in the Meta-model the higher order parameters control the high density behaviour of the EoS instead of the effective nucleon mass and therefore this correlation is not seen. It is however of interest that future $f$-mode detections should be able to put constraints on the higher order nuclear parameters.

\acknowledgments

D. C. and R. M. are grateful to Francesca Gulminelli, Marco Antonelli, Lami Suleiman, Nilaksha Barman, Suprovo Ghosh and Bikram Pradhan for their useful discussions during the project. Their contributions have greatly improved the quality of this work.





\bibliography{ref}

\end{document}